\documentstyle[12pt,epsf,rotate]{article}

\newcommand{\be}{\begin{equation}}
\newcommand{\ee}{\end{equation}}
\newcommand{\bea}{\begin{eqnarray}}
\newcommand{\eea}{\end{eqnarray}}
\newcommand{\bi}{\bibitem}
\newcommand{\nn}{\nonumber}

\newcommand{\complex}{{{\rm I} \kern -.59em {\rm C}}}

\begin{document}


\begin{titlepage}
  \renewcommand{\thefootnote}{\fnsymbol{footnote}}
  \vspace*{-7\baselineskip}
  \begin{flushright}
    \begin{tabular}{l@{}}
       MPI-PhT/96-71\\
       KANAZAWA-96-20\\
       IFUNAM-FT96-11\\
       hep-ph/9703289
    \end{tabular}
  \end{flushright}

  \vskip 0pt plus 0.4fill

  \begin{center}
    \textbf{\LARGE Unification Beyond GUTs: Gauge-Yukawa Unification
    \footnote {Based
    on lectures given by G. Zoupanos in the 
    {\em XXXIV and XXXVI Cracow Schools of Theoretical Particle
    Physics} (Zacopane 1993,1996) and in the {\em 2nd Bruno Pontecorvo
    School on Elementary 
    Particle Physics} (Capri 1996). Partially supported by the E.C.
    projects CHRX-CT93-0319, and ERBFMRXCT960090, the Greek project
    PENED/1170, and the Papiit project IN110296.}
    }

  \end{center}

  \vskip 0pt plus 0.2fill

  \begin{center}
    {\large
    J. Kubo%
    \\}
    \textit{Dept. of Physics,
      Faculty of Science\\
      Kanazawa University\\
      920-11 Kanazawa, Japan}\\
    \vspace{1ex}
    {\large
    M. Mondrag\'on%
    \\}
    \textit{
      Instituto de F\1sica, UNAM\\
      Apdo. Postal 20-364\\
      M\'exico 1000 D.F., M\'exico}\\
      and\\
    {\large
    G. Zoupanos%
    \\} 
    \textit{ 
      Physics Deptartment\\
      Nat. Technical University\\
      157 80 Zografou, Athens, Greece}\\
    
    \vskip 1ex plus 0.3fill


    \vskip 1ex plus 0.7fill

    \textbf{Abstract}
  \end{center}

\noindent
Gauge-Yukawa Unification (GYU) is a renormalization group invariant
functional relation among 
gauge and Yukawa couplings which holds beyond the unification point
in Grand Unified Theories (GUTs).
We present here various models where GYU is obtained by requiring the
principles of 
finiteness and reduction of couplings.  We examine the consequences of
these requirements for the low energy parameters, especially for the
top quark mass. The predictions are such
that they clearly distinguish already GYU from ordinary GUTs. It is
expected that it will be possible to discriminate among the various 
GYUs when more accurate measurements of the top quark
mass are available.
    

  \vskip 0pt plus 2fill

  \setcounter{footnote}{0}

\end{titlepage}

\section{Introduction}

The standard model (SM) is very accurate in describing the
elementary particles and their interactions, but it has a large number
of free parameters whose values are determined only experimentally. 

To reduce the number of free parameters of a theory, and thus render
it more predictive, one is usually led to introduce a symmetry.  Grand
Unified Theories (GUTs) are very good examples of such a procedure
\cite {pati1,gut1,fritzsch1}.  For instance, in the case of minimal $SU(5)$
it was possible to reduce the gauge couplings by one and give a
prediction for one of them.   GUTs can also relate the Yukawa couplings
among themselves, again $SU(5)$ provided an example of this by
predicting the ratio $M_{\tau}/M_b$ \cite{begn} in SM.  Unfortunately,
requiring more gauge symmetry does not seem to help, since additional
complications are introduced due to new degrees of freedom, in the ways
and channels of breaking the symmetry, etc.

A natural extension of the GUT idea is to find a way to relate the
gauge and Yukawa sectors of a theory, that is to achieve Gauge-Yukawa
Unification (GYU).  A symmetry which naturally relates the two sectors
is supersymmetry, in particular $N=2$ supersymmetry.  It turns out,
however, that $N=2$ supersymmetric theories have serious
phenomenological problems due to light mirror fermions.  Also in
superstring theories and in composite models there exist relations
among the gauge and Yukawa couplings, but both kind of theories have
phenomenological problems.

There have been other attempts to relate the gauge and Yukawa
sectors.  One was proposed by Decker, Pestieau, and Veltman
\cite{depe}.  By requiring the absence of quadratic divergences in the
SM, they found a relationship between the squared masses appearing in
the Yukawa 
and in the gauge sectors of the theory.  
A very similar relation is obtained by applying naively in the SM the
general formula derived from demanding spontaneous supersymmetry
breaking via F-terms \cite{fgp-prd90}.
In both cases a prediction for the top quark was possible only when it
was permitted experimentally to neglect the $M_H$ as compared to
$M_{W,Z}$ with the result $M_t=69$ GeV.  Otherwise there is only a
quadratic relation among $M_t$ and $M_H$.

A well known relation among gauge and Yukawa couplings is the
Pendleton-Ross (P-R) {\em infrared} fixed point \cite{pr}.  The P-R
proposal, involving the Yukawa coupling of the top quark $g_t$
and the strong gauge coupling $\alpha_3$, was that the ratio
$\alpha_t/\alpha_3$, where $\alpha_t=g_t^2/4\pi$, has an infrared fixed point.
This assumption predicted
$M_t \sim 100$ GeV.  In addition, it has been shown \cite{zim-pr} that
the P-R 
conjecture is not justified at 
two-loops, since then the ratio $\alpha_t/ \alpha _3$ diverges
in the infrared.

Another interesting conjecture, made by Hill \cite{hill},
is that  $\alpha _t$ itself develops a quasi-infrared fixed point,
leading to the prediction $M_t \sim 280$ GeV. 

The P-R and Hill conjectures have been done in the framework on the SM.
The same conjectures within the minimal supersymmetric SM (MSSM) lead
to the following relations:
\bea
M_t &\simeq & 140 ~{\rm GeV}~\sin \beta ~~~(P-R)\label{3}\\
M_t &\simeq & 200 ~{\rm GeV}~\sin \beta ~~~(Hill)\label{4}
\eea
where $\tan \beta = {v_u/ v_d}$ is the ratio of the two VEV of the
Higgs fields of the MSSM.  We should stress that in  this case there
is no prediction for $M_t$, given that $\sin \beta$ is not fixed from
other considerations.  

In a series of papers \cite{kmz,mondragon2,kmtz,kmtz2,kmz-pert} we
have proposed 
another way to relate the gauge and Yukawa sectors of a theory. It is
based on the fact that 
within the framework of a renormalizable field theory, one can find
renormalization group invariant (RGI) relations among parameters
that can improve the calculability and the predictive power of a
theory.  We have considered models in which the GYU is achieved 
using the principles of reduction of couplings
\cite{cheng1,zim1,kubo1,kubo3,kubo2} and
finiteness
\cite{kmz},\cite{PW}-\cite{model},\cite{LPS}-\cite{piguet},\cite{al}.
These principles, 
which are formulated in 
perturbation theory, are not explicit symmetry principles, although
they might imply symmetries.  The former principle is based on the
existence of RGI relations among couplings, which preserve
perturbative renormalizability.  Similarly, the latter one is based on
the fact that it is possible to find RGI relations among couplings
that keep finiteness in perturbation theory, even to all orders.
Applying these principles one can relate the gauge and Yukawa
couplings without introducing necessarily a symmetry, nevertheless
improving the predictive power of a model.

It is worth noting that the above principles have been applied in
supersymmetric GUTs for reasons that will be transparent in the
following sections.  We should also stress that our conjecture for GYU
is by no means in conflict with the interesting proposals mentioned
before (see also ref.\cite{schrempp1}), but it rather uses all of them,
hopefully in a more successful 
perspective.  For instance, the use of susy GUTs comprises the demand
of the cancellation of quadratic divergences in the SM.  Similarly,
the very interesting conjectures about the infrared fixed points are
generalized in our proposal, since searching for RGI relations among
various couplings corresponds to searching for {\it fixed points} of
the coupled differential equations obeyed by the various couplings of
a theory.

\section{Unification of Couplings by the RGI Method}

Let us next briefly outline the idea of reduction of couplings.  
Any RGI relation among couplings 
(which does not depend on the renormalization
scale $\mu$ explicitly) can be expressed,
in the implicit form $\Phi (g_1,\cdots,g_A) ~=~\mbox{const.}$,
which
has to satisfy the partial differential equation (PDE)
\bea
\mu\,\frac{d \Phi}{d \mu} &=& {\vec \nabla}\cdot {\vec \beta} ~=~ 
\sum_{a=1}^{A} 
\,\beta_{a}\,\frac{\partial \Phi}{\partial g_{a}}~=~0~,
\eea
where $\beta_a$ is the $\beta$-function of $g_a$.
This PDE is equivalent
to a set of ordinary differential equations, 
the so-called reduction equations (REs) \cite{zim1},
\bea
\beta_{g} \,\frac{d g_{a}}{d g} &=&\beta_{a}~,~a=1,\cdots,A~,
\label{redeq}
\eea
where $g$ and $\beta_{g}$ are the primary 
coupling and its $\beta$-function,
and the counting on $a$ does not include $g$.
Since maximally ($A-1$) independent 
RGI ``constraints'' 
in the $A$-dimensional space of couplings
can be imposed by the $\Phi_a$'s, one could in principle
express all the couplings in terms of 
a single coupling $g$.
 The strongest requirement is to demand
 power series solutions to the REs,
\bea
g_{a} &=& \sum_{n=0}\rho_{a}^{(n)}\,g^{2n+1}~,
\label{powerser}
\eea
which formally preserve perturbative renormalizability.
Remarkably, the 
uniqueness of such power series solutions
can be decided already at the one-loop level \cite{zim1}.
To illustrate this, let us assume that the $\beta$-functions
have the form
\bea
\beta_{a} &=&\frac{1}{16 \pi^2}[
\sum_{b,c,d\neq g}\beta^{(1)\,bcd}_{a}g_b g_c g_d+
\sum_{b\neq g}\beta^{(1)\,b}_{a}g_b g^2]+\cdots~,\nn\\
\beta_{g} &=&\frac{1}{16 \pi^2}\beta^{(1)}_{g}g^3+
\cdots~,
\eea
where $\cdots$ stands for higher order terms, and 
$ \beta^{(1)\,bcd}_{a}$'s are symmetric in $
b,c,d$.
 We then assume that
the $\rho_{a}^{(n)}$'s with $n\leq r$
have been uniquely determined. To obtain $\rho_{a}^{(r+1)}$'s,
we insert the power series (\ref{powerser}) into the REs (\ref{redeq})
and collect terms of  
$O(g^{2r+3})$ and find
\bea
\sum_{d\neq g}M(r)_{a}^{d}\,\rho_{d}^{(r+1)} &=&
\mbox{lower order quantities}~,\nn
\eea
where the r.h.s. is known by assumption, and
\bea
M(r)_{a}^{d} &=&3\sum_{b,c\neq g}\,\beta^{(1)\,bcd}_{a}\,\rho_{b}^{(1)}\,
\rho_{c}^{(1)}+\beta^{(1)\,d}_{a}
-(2r+1)\,\beta^{(1)}_{g}\,\delta_{a}^{d}~,\label{M}\\
0 &=&\sum_{b,c,d\neq g}\,\beta^{(1)\,bcd}_{a}\,
\rho_{b}^{(1)}\,\rho_{c}^{(1)}\,\rho_{d}^{(1)}
+\sum_{d\neq g}\beta^{(1)\,d}_{a}\,\rho_{d}^{(1)}
-\beta^{(1)}_{g}\,\rho_{a}^{(1)}~.
\eea
 Therefore,
the $\rho_{a}^{(n)}$'s for all $n > 1$
for a given set of $\rho_{a}^{(1)}$'s can be uniquely determined if
$\det M(n)_{a}^{d} \neq 0$  for all $n \geq 0$.

As it will be clear later by examining specific examples,the various
couplings in supersymmetric theories have easily the same asymptotic
behaviour.  Therefore searching for a power series solution of the
form (\ref{powerser})
to the REs (\ref{redeq}) is justified. This is not the case in
non-supersymmetric theories.

The possibility of coupling unification described in this section  
is without any doubt
attractive because the ``completely reduced'' theory contains 
only one independent coupling, but  it can be
unrealistic. Therefore, one often would like to impose fewer RGI
constraints, and this is the idea of partial reduction \cite{kubo1}.

\section{Partial Reduction in N=1 Supersymmetric Gauge Theories}

Let us consider a chiral, anomaly free,
$N=1$ globally supersymmetric
gauge theory based on a group G with gauge coupling
constant $g$. The
superpotential of the theory is given by
\bea
W&=& \frac{1}{2}\,m_{ij} \,\phi_{i}\,\phi_{j}+
\frac{1}{6}\,C_{ijk} \,\phi_{i}\,\phi_{j}\,\phi_{k}~,
\label{supot}
\eea
where $m_{ij}$ and $C_{ijk}$ are gauge invariant tensors and
the matter field $\phi_{i}$ transforms
according to the irreducible representation  $R_{i}$
of the gauge group $G$. The
renormalization constants associated with the
superpotential (\ref{supot}), assuming that
supersymmetry is preserved, are
\bea
\phi_{i}^{0}&=&(Z^{j}_{i})^{(1/2)}\,\phi_{j}~,~\\
m_{ij}^{0}&=&Z^{i'j'}_{ij}\,m_{i'j'}~,~\\
C_{ijk}^{0}&=&Z^{i'j'k'}_{ijk}\,C_{i'j'k'}~.
\eea
The $N=1$ non-renormalization theorem \cite{nonre} ensures that
there are no mass
and cubic-interaction-term infinities and therefore
\bea
Z_{ijk}^{i'j'k'}\,Z^{1/2\,i''}_{i'}\,Z^{1/2\,j''}_{j'}
\,Z^{1/2\,k''}_{k'}&=&\delta_{(i}^{i''}
\,\delta_{j}^{j''}\delta_{k)}^{k''}~,\nn\\
Z_{ij}^{i'j'}\,Z^{1/2\,i''}_{i'}\,Z^{1/2\,j''}_{j'}
&=&\delta_{(i}^{i''}
\,\delta_{j)}^{j''}~.
\eea
As a result the only surviving possible infinities are
the wave-function renormalization constants
$Z^{j}_{i}$, i.e.,  one infinity
for each field. The one -loop $\beta$-function of the gauge
coupling $g$ is given by \cite{PW}
\bea
\beta^{(1)}_{g}=\frac{d g}{d t} =
\frac{g^3}{16\pi^2}\,[\,\sum_{i}\,l(R_{i})-3\,C_{2}(G)\,]~,
\label{betag}
\eea
where $l(R_{i})$ is the Dynkin index of $R_{i}$ and $C_{2}(G)$
 is the
quadratic Casimir of the adjoint representation of the
gauge group $G$. The $\beta$-functions of
$C_{ijk}$,
by virtue of the non-renormalization theorem, are related to the
anomalous dimension matrix $\gamma_{ij}$ of the matter fields
$\phi_{i}$ as:
\be
\beta_{ijk} =
 \frac{d C_{ijk}}{d t}~=~C_{ijl}\,\gamma^{l}_{k}+
 C_{ikl}\,\gamma^{l}_{j}+
 C_{jkl}\,\gamma^{l}_{i}~.
\label{betay}
\ee
At one-loop level $\gamma_{ij}$ is \cite{PW}
\be
\gamma_{ij}^{(1)}=\frac{1}{32\pi^2}\,[\,
C^{ikl}\,C_{jkl}-2\,g^2\,C_{2}(R_{i})\delta_{ij}\,],
\label{gamay}
\ee
where $C_{2}(R_{i})$ is the quadratic Casimir of the representation
$R_{i}$, and $C^{ijk}=C_{ijk}^{*}$.
Since
dimensional coupling parameters such as masses  and couplings of cubic
scalar field terms do not influence the asymptotic properties 
 of a theory on which we are interested here, it is
sufficient to take into account only the dimensionless supersymmetric
couplings such as $g$ and $C_{ijk}$.
So we neglect the existence of dimensional parameters, and
assume furthermore that
$C_{ijk}$ are real so that $C_{ijk}^2$ always are positive numbers.
For our purposes, it is
convenient to work with the square of the couplings and to
arrange $C_{ijk}$ in such
a way that they are covered by a single index $i~(i=1,\cdots,n)$:
\bea
\alpha &=& \frac{|g|^2}{4\pi}~,~
\alpha_{i} ~=~ \frac{|g_i|^2}{4\pi}~.
\label{alfas}
\eea

The evolution equations of $\alpha$'s in perturbation theory
then take the
form
 \bea
\frac{d\alpha}{d t}&=&\beta~=~ -\beta^{(1)}\alpha^2+\cdots~,\nn\\
\frac{d\alpha_{i}}{d t}&=&\beta_{i}~=~ -\beta^{(1)}_{i}\,\alpha_{i}\,
\alpha+\sum_{j,k}\,\beta^{(1)}_{i,jk}\,\alpha_{j}\,
\alpha_{k}+\cdots~,
\label{eveq}
\eea
where
$\cdots$ denotes the contributions from higher orders, and
$ \beta^{(1)}_{i,jk}=\beta^{(1)}_{i,kj}  $.

Given the set of the evolution equations (\ref{eveq}), we investigate the
asymptotic  properties, as follows. First we
 define \cite{cheng1,zim1}
\bea
\tilde{\alpha}_{i} &\equiv& \frac{\alpha_{i}}{\alpha}~,~i=1,\cdots,n~,
\label{alfat}
\eea
and derive from Eq. (\ref{eveq})
\bea
\alpha \frac{d \tilde{\alpha}_{i}}{d\alpha} &=&
-\tilde{\alpha}_{i}+\frac{\beta_{i}}{\beta}~=~
(\,-1+\frac{\beta^{(1)}_{i}}{\beta^{(1)}}\,)\, \tilde{\alpha}_{i}\nn\\
& &
-\sum_{j,k}\,\frac{\beta^{(1)}_{i,jk}}{\beta^{(1)}}
\,\tilde{\alpha}_{j}\, \tilde{\alpha}_{k}+\sum_{r=2}\,
(\frac{\alpha}{\pi})^{r-1}\,\tilde{\beta}^{(r)}_{i}(\tilde{\alpha})~,
\label{RE}
\eea
where $\tilde{\beta}^{(r)}_{i}(\tilde{\alpha})~(r=2,\cdots)$
are power series of $\tilde{\alpha}$'s and can be computed
from the $r$-th loop $\beta$-functions.
Next we search for fixed points $\rho_{i}$ of Eq. (\ref{alfat}) at $ \alpha
= 0$. To this end, we have to solve
\bea
(\,-1+\frac{\beta ^{(1)}_{i}}{\beta ^{(1)}}\,)\, \rho_{i}
-\sum_{j,k}\frac{\beta ^{(1)}_{i,jk}}{\beta ^{(1)}}
\,\rho_{j}\, \rho_{k}&=&0~,
\label{fixpt}
\eea
and assume that the fixed points have the form
\bea
\rho_{i}&=&0~\mbox{for}~ i=1,\cdots,n'~;~
\rho_{i} ~>0 ~\mbox{for}~i=n'+1,\cdots,n~.
\eea
We then regard $\tilde{\alpha}_{i}$ with $i \leq n'$
 as small
perturbations  to the
undisturbed system which is defined by setting
$\tilde{\alpha}_{i}$  with $i \leq n'$ equal to zero.
As we have seen,
it is possible to verify at the one-loop level \cite{zim1} the
existence of the unique power series solution
\bea
\tilde{\alpha}_{i}&=&\rho_{i}+\sum_{r=2}\rho^{(r)}_{i}\,
\alpha^{r-1}~,~i=n'+1,\cdots,n~
\label{usol}
\eea
of the reduction equations (\ref{RE}) to all orders in the undisturbed
system. 
These are RGI relations among couplings and keep formally
perturbative renormalizability of the undisturbed system.
So in the undisturbed system there is only {\em one independent}
coupling, the primary coupling $\alpha$.

 The small
 perturbations caused by nonvanishing $\tilde{\alpha}_{i}$
 with $i \leq n'$
enter in such a way that the reduced couplings,
i.e., $\tilde{\alpha}_{i}$  with $i > n'$,
become functions not only of
$\alpha$ but also of $\tilde{\alpha}_{i}$
 with $i \leq n'$.
It turned out that, to investigate such partially
reduced systems, it is most convenient to work with the partial
differential equations
\bea
\{~~\tilde{\beta}\,\frac{\partial}{\partial\alpha}
+\sum_{a=1}^{n'}\,
\tilde{\beta_{a}}\,\frac{\partial}{\partial\tilde{\alpha}_{a}}~~\}~
\tilde{\alpha}_{i}(\alpha,\tilde{\alpha})
&=&\tilde{\beta}_{i}(\alpha,\tilde{\alpha})~,\nn\\
\tilde{\beta}_{i(a)}~=~\frac{\beta_{i(a)}}{\alpha^2}
-\frac{\beta}{\alpha^{2}}~\tilde{\alpha}_{i(a)}
&,&
\tilde{\beta}~\equiv~\frac{\beta}{\alpha}~,
\eea
 which are equivalent
to the reduction equations (\ref{RE}), where we let
$a,b$ run from $1$ to $n'$ and $i,j$ from $n'+1$ to $n$
in order to avoid confusion.
We then look for solutions of the form
\bea
\tilde{\alpha}_{i}&=&\rho_{i}+
\sum_{r=2}\,(\frac{\alpha}{\pi})^{r-1}\,f^{(r)}_{i}
(\tilde{\alpha}_{a})~,~i=n'+1,\cdots,n~,
\label{algeq}
\eea
where $ f^{(r)}_{i}(\tilde{\alpha}_{a})$ are supposed to be
power series of
$\tilde{\alpha}_{a}$. This particular type of solution
can be motivated by requiring that in the limit of vanishing
perturbations we obtain the undisturbed
solutions (\ref{usol}) \cite{kubo2,zimmermann3}.
Again it is possible to obtain  the sufficient conditions for
the uniqueness of $ f^{(r)}_{i}$ in terms of the lowest order
coefficients.

\section{The Minimal Asymptotically Free SU(5) Model}

The minimal N=1 supersymmetric SU(5) model \cite{sakai1} is
particularly interesting, 
being the the simplest GUT supported by the LEP data \cite{abf}.  Here we will
consider it as an attractive example of a partially reduced model.  Its
particle content is well defined and has the following transformation
properties under SU(5):
three $({\bf \overline{5}}+{\bf 10})$-
supermultiplets which accommodate three fermion families,
one $({\bf 5}+{\bf \overline{5}})$ to describe the two Higgs
supermultiplets appropriate for electroweak symmetry breaking
and a ${\bf 24}$-supermultiplet required to provide the
spontaneous 
symmetry breaking of $SU(5)$ down to
$SU(3)\times SU(2) \times U(1)$.

Since we are neglecting the dimensional parameters
and the Yukawa couplings of the first two generations,
the superpotential of the model is exactly given by
\begin{equation}
W = \frac{1}{2}\,\,g_{t} {\bf 10}_{3}\,
{\bf 10}_{3}\,H+
g_{b}\, \overline{{\bf 5}}_{3}\,{\bf 10}_{3}\, \overline{H}
+g_{\lambda}\,({\bf 24})^3+
g_{f}\,\overline{H}\,{\bf 24}\, H~,
\end{equation}
where $H, \overline{H}$ are the ${\bf 5},\overline{{\bf 5}}$-
Higgs supermultiplets and we have suppressed the $SU(5)$
indices.
According to the notation introduced in Eq.~(\ref{alfat}), 
Eqs.~(\ref{RE}) become
\bea
\alpha\,\frac{d \tilde{\alpha}_{t}}{d \alpha} &=&
\frac{27}{5}\,\tilde{\alpha}_{t}
-3\,\tilde{\alpha}_{t}^2-\frac{4}{3}\,\tilde{\alpha}_{t}
\tilde{\alpha}_{b}-
\frac{8}{5}\,\tilde{\alpha}_{t}\,\tilde{\alpha}_{f}~,\nn\\
\alpha\,\frac{d \tilde{\alpha}_{b}}{d \alpha} &=&
\frac{23}{5}\,\tilde{\alpha}_{b}
-\frac{10}{3}\,\tilde{\alpha}_{b}^2-\tilde{\alpha}_{b}
\tilde{\alpha}_{t}-
\frac{8}{5}\,\tilde{\alpha}_{b}\,\tilde{\alpha}_{f}~,\nn\\
\alpha\,\frac{d \tilde{\alpha}_{\lambda}}{d \alpha} &=&
9\tilde{\alpha}_{\lambda} -\frac{21}{5}\,\tilde{\alpha}_{\lambda}^2-
\tilde{\alpha}_{\lambda}\,\tilde{\alpha}_{f}~,\nn\\ \alpha\,\frac{d
\tilde{\alpha}_{f}}{d \alpha} &=&
\frac{83}{15}\,\tilde{\alpha}_{f}
-\frac{53}{15}\,\tilde{\alpha}_{f}^2-\tilde{\alpha}_{f}
\tilde{\alpha}_{t}-
\frac{4}{3}\,\tilde{\alpha}_{f}\,\tilde{\alpha}_{b}-
\frac{7}{5}\,\tilde{\alpha}_{f}\,\tilde{\alpha}_{\lambda}~,
\eea
in the one-loop approximation.
Given the above equations describing the evolution of the four
independent couplings $(\alpha_{i}~,~i=t,b,\lambda,f)$,
there exist $2^4=16$ non-degenerate solutions corresponding
to vanishing $\rho$'s as well as non-vanishing ones
given by Eq.~(\ref{algeq}). 
The
possibility to predict the top quark mass depends on a nontrivial
interplay between the vacuum expectation value of the two $SU(2)$
Higgs
doublets involved in the model and the known masses of the third
generation $(m_{b}~,~m_{\tau})$.
It is clear that only the solutions of the form
\be
\rho_{t}~,~\rho_{b}~\neq~0
\label{33}
\ee
can predict the top and bottom quark masses.

There exist exactly  four such solutions.
The first solution is ruled out since it is inconsistent
with Eq.~(\ref{alfas}), and the second one is ruled out
since it does not 
satisfy the criteria to be asymptotically free.
We are left with two asymptotically free solutions, 
which we label 3 and 4 (or AFUT3 and AFUT4, for asymptotically free
unified theory).
According to the criteria of section 3, 
these two solutions
give the possibility to obtain  partial reductions.
To achieve this, we look for solutions \cite{mondragon2} of the form 
Eq.~(\ref{usol}) to both 3 and 4.

We present now the computation of some lower order terms within the
one-loop approximation for the solutions.  For solution 3:
\bea
\tilde{\alpha}_{i} &=& \eta_{i}+ f^{(r_{\lambda}=1)}
_{i}\,\tilde{\alpha}_{\lambda}+f^{(r_{\lambda}=2)}
_{i}\,\tilde{\alpha}_{\lambda}^2+\cdots~\qquad\mbox{for}~i=t,b,f~,
\label{39}
\eea
where
\bea
\eta_{t,b,f} &=&\frac{2533}{2605}~,~
\frac{1491}{2605}~,~\frac{560}{521}~,\nn\\
f^{(r_{\lambda}=1)}_{t,b,f}&\simeq & 0.018~,~0.012~,~-0.131~,\nn\\
f^{(r_{\lambda}=2)}_{t,b,f} &\simeq & 0.005~,~0.004~,~-0.021~,
\label{40}
\eea
For the solution $4$,
\bea
\tilde{\alpha}_{i} &=& \eta_{i}+f^{(r_{f}=1)}_{i}\,
\tilde{\alpha}_{f}+f^{(r_{\lambda}=1)}_{i}\,
\tilde{\alpha}_{\lambda}+f^{(r_{f}=1,r_{\lambda}=1)}_{i}\,
\tilde{\alpha}_{f}\,\tilde{\alpha}_{\lambda}\nn\\
& &+f^{(r_{f}=2)}_{i}\,
\tilde{\alpha}_{f}^2+f^{(r_{\lambda}=2)}_{i}\,
\tilde{\alpha}_{\lambda}^{2}\cdots~~\mbox{for}~i=t,b~,
\label{41}
\eea
where
\bea
 \eta_{t,b}&=&\frac{89}{65}
~,~\frac{63}{65}~,~f^{(r_{\lambda}=1)}_{i}~=~
f^{(r_{\lambda}=2)}_{i}~=~0~,\nn\\
f^{(r_{f}=1)}_{t,b}
&\simeq & -0.258~,
~-0.213~,~f^{(r_{f}=1)}_{t,b}
~\simeq ~ -0.258~,
~-0.213~,\nn\\
f^{(r_{f}=2)}_{t,b}
& \simeq & -0.055~,~ -0.050~,~
f^{(r_{f}=1,r_{\lambda}=1)}_{t,b}
~\simeq ~ -0.021~,~-0.018 ~,
\label{42}
\eea
In the solutions (\ref{39}) and (\ref{41}) we have 
suppressed the contributions
from the Yukawa couplings of the first two generations
because they are negligibly small.

Presumably, both solutions are related;
a numerical analysis on the solutions \cite{mondragon2} suggests that
the solution 
$3$ is a ``boundary''
of $4$. If it is really so,
then there is only one unique reduction solution in the minimal
supersymmetric GUT that provides us with  the possibility of predicting
$\alpha_{t}$. Note furthermore that not only
$\alpha_{t}$ but also $\alpha_{b}$
is predicted in this reduction solution.

Just below the unification scale we would like to obtain the MSSM
$SU(3)\times SU(2)\times U(1)$ and one pair of Higgs doublets, and
assume that all the superpartners 
are degenerate at the supersymmetry breaking scale, where the MSSM
will be broken to the normal SM. 
Then the standard model should  be spontaneously broken down to
$SU(3)\times U(1)_{\rm em}$ due to VEV of the two Higgs
$SU(2)$-doublets contained in the ${\bf 5},\overline{{\bf
5}}$-super-multiplets.

One way to obtain the correct low energy theory is to add to
the Lagrangian soft supersymmetry breaking
terms and to arrange
the mass parameters in the superpotential along with
the soft breaking terms so that
the desired symmetry breaking pattern of the original $SU(5)$
is really the preferred one, all the superpartners are
unobservable at present energies,
there is no contradiction with proton decay,
and so forth.
Then we study 
the evolution of the couplings at two loops
respecting all the boundary conditions at $M_{GUT}$.

\section{Finiteness in N=1 SUSY Gauge Theories}

According to the discussion in Chapter 3, the non-renormalization
theorem ensures there are no extra mass and cubic-interaction-term
renormalizations, implying that the $\beta$-functions of $C_{ijk}$ can
be expressed as linear combinations of the anomalous dimensions
$\gamma_{ij}$ of $\phi^i$.
Therefore, all the one-loop $\beta$-functions of the theory vanish
if $\beta_{g}^{(1)}$  and $\gamma_{ij}^{(1)}$,
given in Eqs. (\ref{betag}) and (\ref{gamay}) respectively, vanish,
i.e.
\begin{equation}
\sum _i \ell (R_i) = 3 C_2(G) \,,
\label{1st}
\end{equation}

\begin{equation}
C^{ikl} C_{jkl} = 2\delta ^i_j g^2  C_2(R_i)\,,
\label{2nd}
\end{equation}

A very interesting result is that the conditions (\ref{1st},\ref{2nd}) are
necessary and sufficient for finiteness at
the two-loop level \cite{PW}.

In case supersymmetry is broken by soft terms, one-loop finiteness of
the soft sector imposes further constraints on it \cite{soft}.  In
addition, the same set of conditions that are sufficient for one-loop
finiteness of the soft breaking terms render the soft sector of they
theory two-loop finite \cite{jj}.

The one- and two-loop finiteness conditions (\ref{1st},\ref{2nd}) restrict
considerably the possible choices of the irreps. $R_i$ for a given
group $G$ as well as the Yukawa couplings in the superpotential
(\ref{supot}).  Note in particular that the finiteness conditions cannot be
applied to the supersymmetric standard model (SSM), since the presence
of a $U(1)$ gauge group is incompatible with the condition
(\ref{1st}), due to $C_2[U(1)]=0$.  This naturally leads to the
expectation that finiteness should be attained at the grand unified
level only, the SSM being just the corresponding, low-energy,
effective theory.

Another important consequence of one- and two-loop finiteness is that
supersymmetry (most probably) can only be broken by soft breaking
terms.  Indeed, due to the unacceptability of gauge singlets, F-type
spontaneous symmetry breaking \cite{raifer} terms are incompatible
with finiteness, as well as D-type \cite{fayet} spontaneous breaking
which requires the existence of a $U(1)$ gauge group.

A natural question to ask is what happens at higher loop orders.  The
answer is contained in a theorem \cite{LPS} which states the necessary
and sufficient conditions to achieve finiteness at all orders.  Before
we discuss the theorem let us make some introductory remarks.  The
finiteness conditions impose relations between gauge and Yukawa
couplings.  To require such relations which render the couplings
mutually dependent at a given renormalization point is trivial.  What
is not trivial is to guarantee that relations leading to a reduction
of the couplings hold at any renormalization point.  As we have seen,
the necessary, but also sufficient, condition for this to happen is to
require that such relations are solutions to the REs
\be
\beta_g {d \lambda_{ijk}\over dg} = \beta_{ijk}
\label{redeq2}
\ee
and hold at all orders.  As we have seen, remarkably the existence of
all-order solutions to (\ref{redeq2}) can be decided at the one-loop
level.

Let us now turn to the all-order finiteness theorem \cite{LPS}, which
states when a $N=1$ supersymmetric gauge theory can become finite to
all orders in the sense of vanishing $\beta$-functions, that is of
physical scale invariance.  It
is based on (a) the structure of the supercurrent in $N=1$ SYM \cite{fz-npb87, pisi-npb196,pisi-book}, and on
(b) the non-renormalization properties of $N=1$ chiral anomalies
\cite{LPS,pisi}. 
Details on the proof can be found in refs. \cite{LPS} and further
discussion in refs.~\cite{pisi,LZ,piguet}.  Here, following mostly
ref.~\cite{piguet} we present a comprehensible sketch of the proof.
 
Consider a $N=1$ supersymmetric gauge theory, with simple Lie group
$G$.  The content of this theory is given at the classical level by
the matter supermultiplets $S_i$, which contain a scalar field
$\phi_i$ and a Weyl spinor $\psi_{ia}$, and the gauge fields $V_a$,
which contain a gauge vector field $A_{\mu}^a$ and a gaugino Weyl
spinor $\lambda^a_{\alpha}$.

Let us first recall certain facts about the theory:

\noindent (1)  A massless $N=1$ supersymmetric theory is invariant 
under a $U(1)$ chiral transformation $R$ under which the various fields 
transform as follows
\be
A'_{\mu}=A_{\mu},~~\lambda '_{\alpha}=\exp({-i\theta})\lambda_{\alpha} 
~~\phi '= \exp({-i{2\over
    3}\theta})\phi,~~\psi_{\alpha}'= \exp({-i{1\over
    3}\theta})\psi_{\alpha},~\cdots
\ee
The corresponding axial Noether current $J^{\mu}_R(x)$ is
\be
J^{\mu}_R(x)=\bar{\lambda}\gamma^{\mu}\gamma^5\lambda + \cdots
\label{noethcurr}
\ee
is conserved classically, while in the quantum case is violated by the
axial anomaly
\be
\partial_{\mu} J^{\mu}_R =
r(\epsilon^{\mu\nu\sigma\rho}F_{\mu\nu}F_{\sigma\rho}+\cdots).
\label{anomaly}
\ee

{}From its known topological origin in ordinary gauge theories
\cite{ag-npb243}, one would expect that the axial vector current
$J^{\mu}_R$ to satisfy the Adler-Bardeen theorem \cite{ab-theo} and
receive corrections only at the one-loop level.  Indeed it has been
shown that the same non-renormalization theorem holds also in
supersymmetric theories \cite{pisi}.  Therefore
\be
r=\hbar \beta_g^{(1)}.
\label{r}
\ee

\noindent (2)  The massless theory we consider is scale invariant at
the classical level and, in general, there is a scale anomaly due to
radiative corrections.  The scale anomaly appears in the trace of the
energy momentum tensor $T_{\mu\nu}$, which is traceless classically.
It has the form
\bea
T^{\mu}_{\mu} &~=~& \beta_g F^{\mu\nu}F_{\mu\nu} +\cdots
\label{Tmm}
\eea

\noindent (3)  Massless, $N=1$ supersymmetric gauge theories are
classically invariant under the supersymmetric extension of the
conformal group -- the superconformal group.  Examining the
superconformal algebra, it can be seen that the subset of
superconformal transformations consisting of translations,
supersymmetry transformations, and axial $R$ transformations is closed
under supersymmetry, i.e. these transformations form a representation
of supersymmetry.  It follows that the conserved currents
corresponding to these transformations make up a supermultiplet
represented by an axial vector superfield called supercurrent
\cite{fz-npb87} $J$,
\be
J \equiv \{ J'^{\mu}_R, ~Q^{\mu}_{\alpha}, ~T^{\mu}_{\nu} , ... \},
\label{J}
\ee
where $J'^{\mu}_R$ is the current associated to R invariance,
$Q^{\mu}_{\alpha}$ is the one associated to supersymmetry invariance,
and $T^{\mu}_{\nu}$ the one associated to translational invariance
(energy-momentum tensor). 

The anomalies of the R current $J'^{\mu}_R$, the trace
anomalies of the 
supersymmetry current, and the energy-momentum tensor, form also
a second supermultiplet, called the supertrace anomaly
\bea
S &= \{ Re~ S, ~Im~ S,~S_{\alpha}\} =\nn\\
& \{T^{\mu}_{\mu},~\partial _{\mu} J'^{\mu}_R,~\sigma^{\mu}_{\alpha
  \dot{\beta}} \bar{Q}^{\dot\beta}_{\mu}~+~\cdots \}
\eea
where $T^{\mu}_{\mu}$ in Eq.(\ref{Tmm}) and
\bea
\partial _{\mu} J'^{\mu}_R &~=~&\beta_g\epsilon^{\mu\nu\sigma\rho}
F_{\mu\nu}F_{\sigma\rho}+\cdots\\ 
\sigma^{\mu}_{\alpha \dot{\beta}} \bar{Q}^{\dot\beta}_{\mu}&~=~&\beta_g
\lambda^{\beta}\sigma^{\mu\nu}_{\alpha\beta}F_{\mu\nu}+\cdots 
\eea

\noindent (4) It is very important to note that 
the Noether current defined in (\ref{noethcurr}) is not the same as the
current associated to R invariance that appears in the
supermultiplet
$J$ in (\ref{J}), but they coincide in the tree approximation. 
So starting from a unique classical Noether current
$J^{\mu}_{R(class)}$,  the Noether
current $J^{\mu}_R$ is defined as the quantum extension of
$J^{\mu}_{R(class)}$ which allows for the
validity of the non-renormalization theorem.  On the other hand
$J'^{\mu}_R$, is defined to belong to the supercurrent $J$,
together with the energy-momentum tensor.  The two requirements
cannot be fulfilled by a single current operator at the same time.

Although the Noether current $J^{\mu}_R$ which obeys (\ref{anomaly})
and the current $J'^{\mu}_R$ 
belonging to the supercurrent multiplet $J$ are not the same, there is a
relation \cite{LPS} between quantities associated with them
\be
r=\beta_g(1+x_g)+\beta_{ijk}x^{ijk}-\gamma_Ar^A
\label{rbeta}
\ee
where $r$ was given in Eq.~(\ref{r}).  The $r^A$ are the
non-renormalized coefficients of 
the anomalies of the Noether currents associated to the chiral
invariances of the superpotential, and --like $r$-- are strictly
one-loop quantities. The $\gamma_A$'s are linear
combinations of the anomalous dimensions of the matter fields, and
$x_g$, and $x^{ijk}$ are radiative correction quantities.
The structure of equality (\ref{rbeta}) is independent of the
renormalization scheme.

One-loop finiteness, i.e. vanishing of the $\beta$-functions at one-loop,
implies that the Yukawa couplings $\lambda_{ijk}$ must be functions of
the gauge coupling $g$. To find a similar condition to all orders it
is necessary and sufficient for the Yukawa couplings to be a formal
power series in $g$, which is solution of the REs (\ref{redeq2}).  

We can now state the theorem for all-order vanishing
$\beta$-functions.
\bigskip

\noindent{\bf Theorem:}

Consider an $N=1$ supersymmetric Yang-Mills theory, with simple gauge
group. If the following conditions are satisfied
\begin{enumerate}
\item There is no gauge anomaly.
\item The gauge $\beta$-function vanishes at one-loop
  \be
  \beta^{(1)}_g = 0 =\sum_i l(R_{i})-3\,C_{2}(G).
  \ee
\item There exist solutions of the form
  \be
  \lambda_{ijk}=\rho_{ijk}g,~\qquad \rho_{ijk}\in\complex
  \label{soltheo}
  \ee
to the  conditions of vanishing one-loop matter fields anomalous dimensions
  \be 
  \gamma^{i~(1)}_j=0=\frac{1}{32\pi^2}~[ ~
  C^{ikl}\,C_{jkl}-2~g^2~C_{2}(R_{i})\delta_{ij} ] . 
  \ee
\item these solutions are isolated and non-degenerate when considered
  as solutions of vanishing one-loop Yukawa $\beta$-functions: 
   \be
   \beta_{ijk}=0.
   \ee
\end{enumerate}
Then, each of the solutions (\ref{soltheo}) can be uniquely extended
to a formal power series in $g$, and the associated super Yang-Mills
models depend on the single coupling constant $g$ with a $\beta$
function which vanishes at all-orders.
\bigskip

It is important to note a few things:
The requirement of isolated and non-degenerate
solutions guarantees the 
existence of a formal power series solution to the reduction
equations.  
The vanishing of the gauge $\beta$-function at one-loop,
$\beta_g^{(1)}$, is equivalent to the 
vanishing of the R current anomaly (\ref{anomaly}).  The vanishing of
the anomalous 
dimensions at one-loop implies the vanishing of the Yukawa couplings
$\beta$-functions at that order.  It also implies the vanishing of the
chiral anomaly coefficients $r^A$.  This last property is a necessary
condition for having $\beta$ functions vanishing at all orders
\footnote{There is an alternative way to find finite theories
  \cite{strassler}.}.  

\bigskip

\noindent{\bf Proof:}

Insert $\beta_{ijk}$ as given by the REs into the
relationship (\ref{rbeta}) between the axial anomalies coefficients and
the $\beta$-functions.  Since these chiral anomalies vanish, we get
for $\beta_g$ an homogeneous equation of the form
\be
0=\beta_g(1+O(\hbar)).
\label{prooftheo}
\ee
The solution of this equation in the sense of a formal power series in
$\hbar$ is $\beta_g=0$, order by order.  Therefore, due to the
REs (\ref{redeq2}), $\beta_{ijk}=0$ too.

Thus we see that finiteness and reduction of couplings are intimately
related. 

\section{Finite $SU(5)$ Model}

As a realistic example of the concepts presented in the previous section
we consider 
a Finite Unified Model Based on $SU(5)$. 
{}From the classification of
theories with vanishing one-loop 
$\beta$ function for the gauge coupling
\cite{HPS}, one can see that
using $SU(5)$ as gauge group there
exist only two candidate models which can 
accommodate three fermion
generations. These models contain the chiral supermutiplets
${\bf 5}~,~\overline{\bf 5}~,~{\bf 10}~,
~\overline{\bf 5}~,~{\bf 24}$
with the multiplicities $(6,9,4,1,0)$ and
 $(4,7,3,0,1)$, respectively.
Only the second one contains a ${\bf 24}$-plet which can be used
for spontaneous symmetry breaking (SSB) of $SU(5)$ down
to $SU(3)\times SU(2) \times U(1)$. (For the first model
one has to incorporate another way, such as the Wilson flux
breaking to achieve the desired SSB of $SU(5)$ \cite{kmz}).
Therefore,  we would like to concentrate only on the second model.

To simplify the situation, we neglect the intergenerational
mixing among the lepton and quark supermultiplets and consider
the following $SU(5)$ invariant cubic
superpotential for the (second)
model:
\bea
W &=& \sum_{i=1}^{3}\sum_{\alpha=1}^{4}\,[~\frac{1}{2}g_{i\alpha}^{u}
\,{\bf 10}_i
{\bf 10}_i H_{\alpha}+
+g_{i\alpha}^{d}\,{\bf 10}_i \overline{\bf 5}_{i}\,
\overline{H}_{\alpha}~] \nn\\
 & & +\sum_{\alpha=1}^{4}g_{\alpha}^{f}\,H_{\alpha}\, 
{\bf 24}\,\overline{H}_{\alpha}+
\frac{g^{\lambda}}{3}\,({\bf 24})^3~,~
\mbox{with}~~g_{i \alpha}^{u,d}=0~\mbox{for}~i\neq \alpha~,
\eea
where the ${\bf 10}_{i}$'s
and $\overline{\bf 5}_{i}$'s are the usual
three generations, and the four
$({\bf 5}+ \overline{\bf 5})$ Higgses are denoted by
 $H_{\alpha}~,~\overline{H}_{\alpha} $.
The superpotential is not the most general one, but
by virtue of the non-renormalization theorem,
this does not contradict the philosophy of 
the coupling unification by the reduction 
method (a RG invariant fine tuning is a solution
of the reduction equation). In the case at hand,
however, one can
find a discrete symmetry that can be imposed
on the most general cubic superpotential to arrive at the
non-intergenerational mixing \cite{kmz}.  This is given in Table 1.

\begin{table*}[t]
\caption{The charges of the $Z_7\times Z_3$ symmetry}
\label{tbl}
$$
\begin{tabular}{|c|c|c|c|c|c|c|c|c|c|c|}
\hline
& ${\bf 10}_1$ & ${\bf 10}_2$ &
${\bf 10}_3$ &
 $\bar {\bf 5}_1$ & $\bar {\bf 5}_2
$ & $\bar {\bf 5}_3$& $H_1$ &
  $H_2$ & $H_3$ & $H_4$ \\ \hline
$Z_7$ &1 & 2  & 4 &  4  &
1  &  2  & 5&  3 & 6 & 0\\ \hline
$Z_3$ &1 &2   &0  & 0 &0
&0 &1 & 2 & 0 & 0\\ \hline
\end{tabular}
$$
\end{table*}

Given the superpotential $W$,
we  can  compute the $\beta$ functions of the model. 
We denote the gauge coupling  by $g$
(with the vanishing one-loop $\beta$ function), and 
our  normalization of the $\beta$ functions
is as usual, i.e., 
$$d g_{i}/d \ln \mu ~=~
\beta^{(1)}_{i}/16 \pi^2+O(g^5),$$
where $\mu$ is the renormalization
scale. We find: 
\bea
\beta^{(1)}_{g} &=& 0~,\nn\\
\beta^{u(1)}_{i\alpha} &=& \frac{1}{16\pi^2}\,
[\,-\frac{96}{5}\,g^2+
6\,\sum_{\beta=1}^{4}(g_{i\beta}^{u})^{2}+
3\,\sum_{j=1}^{3}(g_{j\alpha}^{u})^{2}
+\frac{24}{5}\,(g^{f}_{\alpha})^{2}\nn\\
&&+4\,\sum_{\beta=1}^{4}(g_{i\beta}^{d})^{2}
\,]\,g_{i\alpha}^{u}~,\nn\\
\beta^{d(1)}_{i\alpha} &=& \frac{1}{16\pi^2}\,
[\,-\frac{84}{5}\,g^2+
3\,\sum_{\beta=1}^{4}(g_{i\beta}^{u})^{2}
+\frac{24}{5}\,(g^{f}_{\alpha})^{2}+
4\,\sum_{j=1}^{3}(g_{j\alpha}^{d})^{2}\nn\\
&&+6\,\sum_{\beta =1}^{4}(g_{i\beta}^{d})^{2}\,]\,g_{i\alpha}^{d}~,\\
\beta^{\lambda(1)} &=& \frac{1}{16\pi^2}\,
[\,-30\,g^2+\frac{63}{5}\,(g^{\lambda})^2+
3\,\,\sum_{\alpha =1}^{4}(g_{ \alpha}^{f})^{2}
\,]\,g^{\lambda}~,\nn\\
\beta^{f(1)}_{\alpha} &=& \frac{1}{16\pi^2}\,
[\,-\frac{98}{5}\,g^2+3\,\sum_{i=1}^{3}(g_{i \alpha}^{u})^{2}
+4\,\sum_{i=1}^{3}(g_{i \alpha}^{d})^{2}
+\frac{48}{5}\,(g^{f}_{\alpha})^{2}\nn\\
&&+\sum_{\beta=1}^{4}(g_{\beta}^{f})^{2}
+\frac{21}{5}\,(g^{\lambda})^{2}
\,]\,g_{\alpha}^{f}~.\nn
\eea
We then regard the gauge coupling $g$ as the primary
coupling and solve the reduction equations (\ref{redeq}) with
the power series ansatz. One finds that the power series,
\bea
(g_{i i}^{u})^2 &=&\frac{8}{5}g^2+\dots~,
~(g_{i i}^{d})^2 =\frac{6}{5}g^2+\dots~,~
(g^{\lambda})^2=\frac{15}{7}g^2+\dots~,\nn\\
(g^{f}_{4})^2 &=& g^2~,~(g^{f}_{\alpha})^2=0+\dots~~(\alpha=1,2,3)~,
\label{fut-bc}
\eea
exists uniquely,
where $\dots$ indicates higher order terms and
all the other couplings have to vanish.
As we have done in the previous section, we can easily 
verify that the higher order terms can be uniquely
computed.

 Consequently, all the one-loop $\beta$ functions of the theory vanish.
Moreover, all the one-loop anomalous dimensions for the chiral
supermultiplets,
\bea
\gamma^{(1)}_{{\bf 10}i} &=& \frac{1}{16\pi^2}\,
[\,-\frac{36}{5}\,g^2+
3\,\sum_{\beta=1}^{4}(g_{i\beta}^{u})^{2}+
2\,\sum_{\beta=1}^{4}(g_{i\beta}^{d})^{2}
\,]~,\nn\\
\gamma^{(1)}_{\overline{{\bf 5}}i} &=& \frac{1}{16\pi^2}\,
[\,-\frac{24}{5}\,g^2+
4\,\sum_{\beta =1}^{4}(g_{i\beta}^{d})^{2}\,]~,\nn\\
\gamma^{(1)}_{H_{\alpha}} &=& \frac{1}{16\pi^2}\,
[\,-24\,g^2+
3\,\,\sum_{i =1}^{3}(g_{i\alpha}^{u})^{2}+
\frac{24}{5}(g_{\alpha}^{f})^2\,]~,\\
\gamma^{(1)}_{\overline{H}_{\alpha}} &=& \frac{1}{16\pi^2}\,
[\,-24\,g^2+
4\,\,\sum_{i =1}^{3}(g_{i\alpha}^{d})^{2}+
\frac{24}{5}(g_{\alpha}^{f})^2\,]~,\nn\\
\gamma^{(1)}_{{\bf 24}} &=& \frac{1}{16\pi^2}\,
[\,-\frac{10}{5}\,g^2+
+\sum_{\alpha=1}^{4}(g_{\alpha}^{f})^{2}
+\frac{21}{5}\,(g^{\lambda})^{2}\,]~,\nn
\eea
also vanish in the reduced system.
As it has already been mentioned before, these conditions  are
necessary and sufficient for finiteness to all orders in perturbation
theory. 

In most of the previous studies of
the present model \cite{model1,model}, however,
the complete reduction of the Yukawa couplings,
which is necessary for all-order-finiteness,
was ignored.  They have used the freedom
offered by the degeneracy in the one- and two-loop
approximations in order to make
specific ans{\" a}tze that could lead to phenomenologically acceptable
predictions.
In the above model, we found a diagonal solution for the Yukawa
couplings, with each family coupled to a different Higgs.
However, we may use the fact that mass terms
do not influence the RG functions in a certain
class of renormalization schemes, and introduce
appropriate mass terms that permit us to perform a rotation in the Higgs
sector such that only one pair of Higgs doublets, coupled to
the third family, remains light and acquires a
non-vanishing VEV \cite{model}. 
Note that the effective coupling of the Higgs doublets
to the first family after
the rotation is very small avoiding in this way a potential problem
with the proton lifetime \cite{proton}.
Thus, effectively,
we have at low energies the Minimal Supersymmetric Standard Model
(MSSM) with
only one pair of Higgs doublets
satisfying the boundary conditions at $M_{\rm GUT}$
\be
 g_{t}^{2}  = \frac{8}{5} g^2+O(g^4)~,~
 g_{b}^{2}=g_{\tau}^{2}=\frac{6}{5} g^2+O(g^4)~,
\ee
where $g_i$ ($i=t, b, \tau$) are the top, bottom
and tau Yukawa couplings
of the MSSM, and the other Yukawa couplings 
should be regarded as free.

Adding soft
breaking terms (which are supposed not to influence the
$\beta$ functions beyond $M_{\rm GUT}$),
we can obtain supersymmetry breaking.
The conditions on the soft breaking terms to preserve
one-loop finiteness have been given already some time ago
\cite{soft}. 
Recently, the same problem
in two-loop orders has been addressed \cite{jj}.
It is an open problem whether there exists a suitable set of conditions
on the soft terms for all-loop finiteness.

\section{Predictions of Low Energy Parameters}

In this section we will refine 
the predictions of the AFUT and FUT models, taking into account certain
corrections and we will compare them with the experimental data.

As mentioned before, at low energies we want the MSSM, with one pair
of Higgs doublets, 
and we will assume that at the supersymmetry breaking scale all the
superpartners are degenerate.

Since the gauge symmetry is spontaneously broken
below $M_{\rm GUT}$, the finiteness conditions in the case of the FUT
model do not restrict the renormalization property at low energies, and
all it remains is a boundary condition on the
gauge and Yukawa couplings at $M_{\rm GUT}$, i.e., Eq. (\ref{fut-bc}).
Clearly the same holds also in the AFUT models.
So we examine the evolution of these couplings according
to their renormalization group equations at two-loops with
the corresponding boundary conditions at $M_{\rm GUT}$.

Below $M_{\rm GUT}$ their evolution is assumed to be
governed by the MSSM. We further assume a unique threshold
$M_{\rm SUSY}$ for all superpartners of the MSSM so that
below $M_{\rm SUSY}$ the SM is the correct effective theory.
We  recall that
$\tan\beta$ is usually determined in the Higgs sector, which however
strongly depends on the supersymmetry breaking terms.
Here we avoid this by using the tau mass $M_{\tau}$ 
as input, which means that we partly fix the Higgs sector
indirectly.
That is, assuming that
\be
M_Z \ll M_{t} \ll M_{\rm SUSY}~,
\ee
we require the matching condition at $M_{\rm SUSY}$ \cite{barger},
\bea
\alpha_{t}^{\rm SM} 
&=&\alpha_{t}\,\sin^2 \beta~,~
\alpha_{b}^{\rm SM}
~ =~ \alpha_{b}\,\cos^2 \beta~,
~\alpha_{\tau}^{\rm SM}
~=~\alpha_{\tau}\,\cos^2 \beta~,\nn\\
\alpha_{\lambda}&=&
\frac{1}{4}(\frac{3}{5}\alpha_{1}
+\alpha_2)\,\cos^2 2\beta~,
\label{match}
\eea
to be satisfied,
where $\alpha_{i}^{\rm SM}~(i=t,b,\tau)$ are
the SM Yukawa couplings and $\alpha_{\lambda}$ is the Higgs coupling.
 The MSSM threshold corrections
to this matching condition \cite{hall1,wright1} will be discussed
later.
 This is our definition of $\tan\beta$, and Eq.~(\ref{match})
 fixes $\tan\beta$, because with a given set of the input
parameters \cite{pdg}, 
\bea
M_{\tau} &=&1.777 ~\mbox{GeV}~,~M_Z=91.188 ~\mbox{GeV}~,
\label{mtau}
\eea
with \cite{pokorski1}
\bea
\alpha_{\rm EM}^{-1}(M_{Z})&=&127.9
+\frac{8}{9\pi}\,\log\frac{M_t}{M_Z} ~,\nn\\
\sin^{2} \theta_{\rm W}(M_{Z})&=&0.2319
-3.03\times 10^{-5}T-8.4\times 10^{-8}T^2~,\\
T &= &M_t /[\mbox{GeV}] -165~,\nn
\label{aem}
\eea
the matching condition (\ref{match}) and the GYU
boundary condition at $M_{\rm GUT}$ can be satisfied only for a specific
value of $\tan\beta$. Here  $M_{\tau},M_t, M_Z$
are pole masses, and the couplings are defined in the 
$\overline{\mbox{MS}}$ scheme with six flavors.
The translation from a Yukawa coupling
into the corresponding mass follows according to
\bea
m_i&=&\frac{1}{\sqrt{2}}g_i(\mu)\,v(\mu)~,~i=t,b,\tau ~~
\mbox{with} ~~v(M_Z)=246.22~\mbox{GeV}~,
\label{mass-yuk}
\eea
where $m_i(\mu)$'s are the running masses satisfying
the respective evolution equation of two-loop order.
The pole masses can be calculated from the
running ones of course. For the top mass, we use \cite{barger,hall1}
\bea
M_{t} &=&m_{t}(M_t)\,[\,1+
\frac{4}{3}\frac{\alpha_3(M_t)}{\pi}+
10.95\,(\frac{\alpha_3(M_t)}{\pi})^2+k_t 
\frac{\alpha_t(M_t)}{\pi}\,]~,
\label{top-mass}
\eea
where  $k_t \simeq -0.3$ for the range of parameters
we are concerned with in this paper \cite{hall1}.
Note that both sides of Eq.~(\ref{top-mass}) contain $M_t$ so that
$M_t$ is defined only implicitly.
Therefore, its determination requires an iteration method.
As for the tau and bottom masses, we assume that
$m_{\tau}(\mu)$ and $m_b(\mu)$ for $\mu \leq M_Z$
satisfy the evolution equation governed by
the $SU(3)_{\rm C}\times U(1)_{\rm EM}$ theory 
with five flavors and use
\bea
M_{b}&=&m_b(M_b)\,[\,1+
\frac{4}{3}\frac{\alpha_{3(5{\rm f})}(M_b)}{\pi}+
12.4\,(\frac{\alpha_{3(5{\rm f})}(M_b)}{\pi})^2\,]~,\nn\\
M_{\tau}&=&m_{\tau}(M_{\tau})\,[\,1+
\frac{\alpha_{\rm EM (5f)}(M_{\tau})}{\pi}\,]~,
\eea
where the experimental value of $m_b(M_b)$ is
$(4.1-4.5)$ GeV \cite{pdg}.
The couplings with five flavors entered in Eq. (30)
$\alpha_{3(5{\rm f})}$ and $\alpha_{\rm EM (5f)}$
are related to $\alpha_{3}$ and $\alpha_{\rm EM}$ by
\bea
\alpha_{3(5{\rm f})}^{-1}(M_Z) &= &\alpha_{3}^{-1}(M_Z)
-\frac{1}{3\pi}\,\ln \frac{M_t}{M_Z} ~,\nn\\
\alpha_{\rm EM (5f)}^{-1}(M_Z) &= & \alpha_{\rm EM}^{-1}(M_Z)-
\frac{8}{9\pi}\,\ln \frac{M_t}{M_Z}~.
\eea
Using the input values given in eqs.~(\ref{mtau}) and (\ref{aem}), we find
\bea
m_{\tau}(M_{\tau})&=&1.771~\mbox{GeV}~,
m_{\tau}(M_{Z})=1.746~\mbox{GeV}~,\nn\\
\alpha_{\rm EM (5f)}^{-1}(M_{\tau})&=&133.7~,
\eea
and from Eq.~(\ref{mass-yuk}) we 
obtain 
\bea
\alpha_{\tau}^{\rm SM}(M_Z)&=&\frac{g_{\tau}^{2}}{4\pi}
=8.005\times 10^{-6}~,
\eea
which we use as an input parameter instead of $M_{\tau}$.

The matching condition (\ref{match})  suffers from the threshold 
corrections coming from the MSSM superpartners:
\bea
\alpha_{i}^{\rm SM} \to 
\alpha_{i}^{\rm SM}(1+\Delta_{i}^{\rm SUSY})~,~i=1,2,\dots,\tau~,
\eea
It was shown that these threshold effects to
the  gauge couplings can
be effectively parametrized by just one energy  scale 
\cite{langacker1}. 
Accordingly, we can identify our $M_{\rm SUSY}$ with that defined
in ref.\cite{langacker1}.  This ensures that there are no further
one-loop threshold corrections  to $\alpha_3(M_Z)$ when we
calculate it as a function of  $\alpha_{\rm EM}(M_Z)$ and
$\sin^2\theta_W(M_Z)$.

The same scale $M_{\rm SUSY}$
does not describe  threshold corrections to the Yukawa
couplings,
and they could cause large corrections to the fermion mass
 prediction \cite{hall1,wright1} \footnote{It is
possible to compute the MSSM correction to $M_t$ directly, i.e., 
without constructing an effective theory below $M_{\rm SUSY}$.
In this approach, too,  large corrections have  been reported
\cite{polonsky1}. In the present paper, evidently,  we are following
the effective theory approach as 
e.g. refs. \cite{hall1,wright1}.}.
For $m_b$, for instance, the correction  can be as large as 50\%
for very large values  of $\tan\beta$,
especially in models with radiative 
gauge symmetry breaking and with supersymmetry softly broken by 
the universal breaking terms. As we will see, the $SU(5)$-FUT and AFUT
models predict (with these corrections suppressed) values
for the bottom quark mass that are 
rather close to the experimentally allowed region 
so that there is room only for small corrections.
Consequently, if we want to break
$SU(2) \times U(1)$ gauge 
symmetry radiatively, the models favor
non-universal soft breaking terms \cite{borzumati1}.

To get an idea about the magnitude of the correction, 
we consider
the case that all the superpartners 
have the same mass $M_{\rm SUSY}=500$ GeV with
$M_{\rm SUSY} \gg \mu_H$ and $\tan\beta \geq 50$.
Using  $\Delta$'s 
given in  ref. \cite{wright1}, we find that
the MSSM correction to the $M_t$ prediction
is $\sim -1$ \% for this case.
Comparing with the 
results of \cite{wright1,polonsky1}, 
this may appear to be underestimated for other cases.
Note, however, that there is a nontrivial interplay among the 
corrections between the $M_t$ and $M_b$ predictions
for a given GYU boundary condition at $M_{\rm GUT}$
and the fixed pole tau mass, which has not been taken into
account in refs. \cite{wright1,polonsky1}. 
In the following discussion, therefore,
we regard the MSSM threshold correction to
the $M_t$ prediction as unknown and denote it by
\be
\delta^{\rm MSSM} M_t~.
\ee

  In the case of the AFUT
models, the non-observation of proton decay favours  a solution close
to AFUT3.    

In table 2 we present
the predictions for $M_t$ for various $M_{\rm SUSY}$, in
the case of the FUT model.

\begin{table}
\caption{The predictions 
for different $M_{\rm SUSY}$ for FUT}
\begin{center}
\begin{tabular}{|c|c|c|c|c|c|}
\hline
$M_{\rm SUSY}$ [GeV]   &$\alpha_{3}(M_Z)$ &
$\tan \beta$  &  $M_{\rm GUT}$ [GeV] 
 & $m_b (M_{b}) $ [GeV]& $M_{t}$ [GeV]
\\ \hline
$300$ & $0.123 $  &$54.2 $  & $2.08\times 10^{16}$
 & $4.54$  & 183.5\\ \hline
$500$ & $0.122 $  &$54.3 $  & $1.77\times 10^{16}$
 & $4.54$  & 184.0 \\ \hline
$10^3$ & $0.120 $  &$54.4 $  & $1.42\times 10^{16}$
 & $4.54$  & 184.4 \\ \hline
\end{tabular}
\end{center}
\end{table}

\noindent
As we can see from the table,  only negative MSSM
corrections of at most $\sim 10$ \% to $m_b(M_b)$ 
is allowed ( $m_{b}^{\rm exp}(M_b)=
(4.1-4.5)$ GeV), implying that FUT
favors  non-universal soft symmetry breaking terms as announced.
The predicted $M_t$ values are well below the 
infrared value \cite{hill1},
for instance
$194$ GeV for $M_{\rm SUSY}=500$ GeV, 
so that the $M_t$ prediction
must be sensitive against the change of the boundary condition.

We recall that if one includes
the threshold effects of superheavy particles \cite{threshold},
the GUT scale $M_{\rm GUT}$ at which $\alpha_1$ and $\alpha_2$
are supposed to meet is related to
the mass of the superheavy
$SU(3)_C $-triplet  Higgs supermultiplets contained 
in $H_{\alpha}$ and $\overline{H}_{\alpha}$.
These  effects  have therefore
influence on the GYU boundary conditions.

\begin{table}
\caption{The predictions for the AFUT model}\label{table-afut} 
\begin{center}
\vspace{0.4 cm}
\begin{tabular}{|c|c|c|c|c|c|}
\hline
$m_{\rm SUSY}$ [GeV]&
$\alpha_{3}(M_Z)$ &
$\tan \beta$  &  $M_{\rm GUT}$ [GeV]
 & $m_{b} $ [GeV]& $m_{t}$ [GeV]
\\ \hline
$300$ &  $0.120$  & $47.7$ & $1.8\times10^{16}$
  & $5.4$  & $179.7 $  \\ \hline
$500$ &  $0.118$  & $47.7$ & $1.39\times10^{16}$
  & $5.3$  & $178.9$  \\ \hline
\end{tabular}
\end{center}
\end{table}

In Table 3 we present the predictions for the AFUT viable model (AFUT3).
For these model the corrections mentioned above
have been calculated \cite{kmoz2} and are of the order of $\leq 2 \%$.
The threshold effects of the superheavy particles were estimated to be
of the same order as in the gauge sector, which leads to an uncertainty
of $\sim \pm 0.4$ GeV in $M_t$.
The structure of the threshold effects in FUT
 is involved, but
they are not arbitrary and probably determinable to a certain
extent, because the mixing of the superheavy Higgses
is strongly dictated by the fermion mass matrix of the MSSM.
To bring these threshold effects under control is
challenging.
Here we assume that the magnitude of these effects is
$\sim \pm 4$  GeV in $M_t$, which is estimated by comparing the
minimal GYU model based on $SU(5)$ \cite{kmoz2}. 

Thus, for the FUT model the prediction for $M_t$ \cite{kmoz2} will be
\be
M_t =(183+\delta^{\rm MSSM} M_t\pm 5) ~~\mbox{GeV}~,
\ee
where the finite corrections coming from the conversion
from the dimensional reduction scheme to
the ordinary $\overline{\mbox{MS}}$ 
in the gauge sector \cite{anton} are included, and
those in the Yukawa sector are included as an uncertainty
of $\sim\pm 1$ GeV.
The MSSM  threshold
correction 
is  denoted $\delta^{\rm MSSM} M_t$
which has been discussed in the previous section.

In the case of the AFUT model the prediction is \cite{kmoz2}
\be
M_t =(181+\delta^{\rm MSSM} M_t\pm 3) ~~\mbox{GeV}~.
\ee

Comparing the $M_t$ prediction above  with the most recent experimental
values \cite{top}, 
\bea
M_{top} &=& 176.8 ~\pm 4.4_{stat}~\pm 4.8_{syst}~~{\rm GeV}~~~{\rm CDF}\nn\\
M_{top} &=& 169.0 ~\pm 8.0_{stat}~\pm 8.0_{syst}~~{\rm GeV}~~~{\rm D0}
\label{mtop-exp}
\eea
 we see it is consistent with the experimental data.

\begin{figure}[t]
  \begin{center}
    \leavevmode
    \rotate[r]{
      \mbox{
        \epsfysize=11cm
        \epsffile{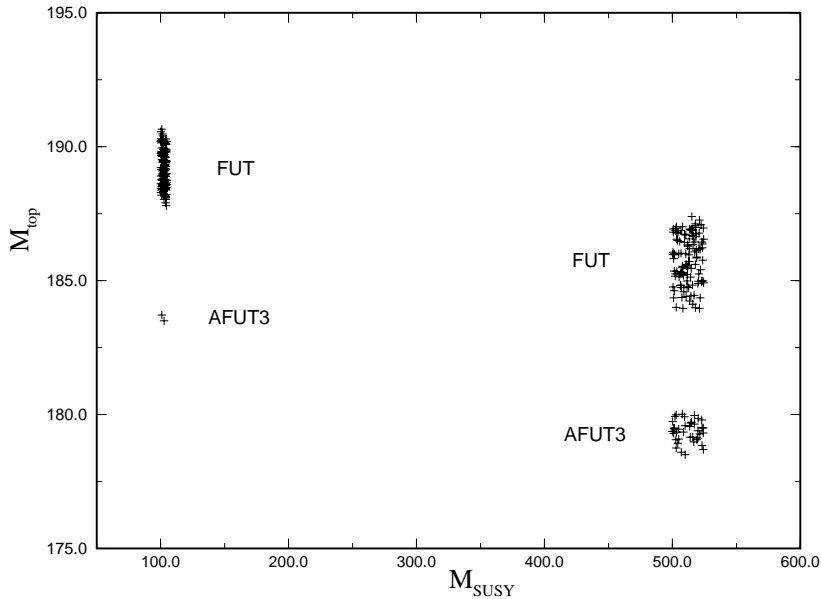}
        }
      }
  \end{center}
  \caption[\ ]{ $M_t$ predictions of $SU(5)$ FUT and AFUT3 models, for
    given $M_{SUSY}$ around 100 and 500 GeV. For
    the FUT model $\tilde{\alpha}_t=1.6$, $\tilde{\alpha}_b=1.2$, and for
    AFUT3 $\tilde{\alpha}_t=0.97$, $\tilde{\alpha}_b=0.57$.}
  \label{fig:4}
\end{figure}

It is interesting to note that
the consistency of the finiteness hypothesis
is closely related to the fine structure of supersymmetry breaking
and also to the Higgs sector, because
these superpartner corrections  to $m_b$ can be kept small
for appropriate 
supersymmetric spectrum characterized by very heavy squarks 
and/or small  $\mu_H$ describing the mixing of the two 
Higgs doublets in the superpotential 
\footnote{The solution with small $\mu_H$ 
is favored by the experimental data and cosmological constraints
\cite{borzumati1}. The sign of this correction 
is determined by the relative sign of 
$\mu_H$ and the gluino mass parameter and is correlated 
with the chargino exchange contribution 
to the $b \to s \gamma$ decay \cite{hall1}. 
The later has the same sign as the Standard Model and the charged 
Higgs contributions when the supersymmetric corrections to $m_b$ are 
negative.}.

The predictions for $M_t$ versus $M_{SUSY}$ for the two sets of
boundary conditions given above (AFUT3 and AFUT4) together with the
corresponding predictions of the FUT model, are given in Figure 1.
In a  recent study \cite{kmoz2}, we have 
considered the proton decay constraint \cite{HMY-npb402}
to further reduce the parameter space of the model.
It has been found that the model consistent with the 
non-observation of the proton decay
should be very close to AFUT3, implying a better
possibility to discriminate between the FUT and AFUT models,
as one can see from Figure \ref{fig:4}.

\section{Asymptotically Non-Free Supersymmetric Pati-Salam Model}

We present now a model where the reduction of couplings is applied,
but that does not have a single gauge group, but a product of simple
groups. 
In order for the RGI method for the gauge coupling
unification to  work,
the gauge couplings should 
have the same asymptotic behavior.
Note that this common behavior is absent
in the standard model with three families.
A way to achieve a common asymptotic behavior of all the
different gauge couplings is to embed 
$SU(3)_{C}\times SU(2)_{L}\times U(1)_{Y}$ to some
non-abelian gauge group, as it was done in the previous sections.
However, in this case still a major r\^ole in the
GYU is due to the group theoretical aspects of the covering GUT. Here
we would like  to examine the power of RGI method by considering
theories without covering GUTs.
We found \cite{kmtz}
that the minimal phenomenologically viable model is based on the gauge
group of Pati and 
Salam \cite{pati1}-- ${\cal G}_{\rm PS}\equiv 
SU(4)\times SU(2)_{R}\times
SU(2)_{L}$.
We recall that $N=1$ supersymmetric  models based on this
gauge group have been studied with renewed interest because they could
in principle be derived from superstring \cite{anton1}.

In our supersymmetric, Gauge-Yukawa unified model
based on $ {\cal G}_{\rm PS}$ \cite{kmtz}, three generations of
quarks and leptons  are accommodated by six chiral supermultiplets, three
in $({\bf 4},{\bf 2},{\bf 1})$ and three  $({\bf \overline{4}},{\bf
1},{\bf 2})$, which we denote by $\Psi^{(I)\mu~ i_R}$ and $
\overline{\Psi}_{\mu}^{(I) i_L}$. ($I$ runs over the three generations,
and
$\mu,\nu~(=1,2,3,4)$ are the $SU(4)$ indices while 
$i_R~,~i_L~(=1,2)$ 
stand for the
$SU(2)_{L,R}$ indices.) 
The Higgs supermultiplets 
in $({\bf 4},{\bf 2},{\bf 1})$,
$({\bf \overline{4}},{\bf 2},{\bf 1})$
and  $({\bf 15},{\bf 1},{\bf 1})$ are denoted by 
$ H^{\mu ~i_R}~,~
\overline{H}_{\mu ~i_R} $ and $\Sigma^{\mu}_{\nu}$, respectively. They
 are responsible for the spontaneous
symmetry breaking (SSB) of $SU(4)\times SU(2)_{R}$ down 
to $SU(3)_{C}\times U(1)_{Y}$.
The SSB of $U(1)_{Y}\times
SU(2)_{L}$ is then achieved by the nonzero VEV of
$h_{i_R i_L}$ which is in $({\bf 1},{\bf 2},{\bf 2})$. In addition to
these Higgs supermultiplets, we introduce $G^{\mu}_{\nu~i_R i_L}~
({\bf 15},{\bf 2},{\bf 2})~,
~\phi~({\bf 1},{\bf 1},{\bf 1})$ and 
$\Sigma^{' \mu}_{\nu}~({\bf 15},{\bf 1},{\bf 1})$.
The $G^{\mu}_{\nu~i_R i_L}$ is introduced to realize 
the $SU(4)\times SU(2)_{R}\times
SU(2)_{L}$ version of the Georgi-Jarlskog type 
ansatz \cite{georgi4} for
the mass matrix of leptons and quarks while $\phi$ 
is supposed to mix with the right-handed neutrino
supermultiplets at a high energy scale.
With these things in mind, we write down
the superpotential of the model
$W$, which is the sum of the following superpotentials:
\bea
W_{Y} &=&\sum_{I,J=1}^{3}g_{IJ}\,\overline{\Psi}^{(I) i_R}_{\mu} 
\,\Psi^{(J)\mu~ i_L}~h_{i_R i_L}~,~\nn\\
W_{GJ} &=& g_{GJ}\,
\overline{\Psi}^{(2)i_R}_{\mu}\,
G^{\mu}_{\nu~i_R j_L}\,\Psi^{(2)\nu~ j_L}~,\nn\\
W_{NM} &=&
\sum_{I=1,2,3}\,g_{I\phi}~\epsilon_{i_R j_R}\,\overline{\Psi}^{(I)
i_R}_{\mu} ~H^{\mu ~j_R}\,\phi~,\nn\\
W_{SB} &=&
g_{H}\,\overline{H}_{\mu~ i_R}\,
\Sigma^{\mu}_{\nu}\,H^{\nu ~i_R}+\frac{g_{\Sigma}}{3}\,
\mbox{Tr}~[~\Sigma^3~]+
\frac{g_{\Sigma '}}{2}\,\mbox{Tr}~[~(\Sigma ')^2\,\Sigma~]~,\nn\\
W_{TDS} &=&
\frac{g_{G}}{2}\,\epsilon^{i_R j_R}\epsilon^{i_L j_L}\,\mbox{Tr}~
[~G_{i_R i_L}\,
\Sigma\,G_{j_R j_L}~]~,\nn\\
W_{M}&=&m_{h}\,h^2+m_{G}\,G^2+m_{\phi}\,
\phi^2+m_{H}\,\overline{H}\,H+
m_{\Sigma}\,\Sigma^2+
m_{\Sigma '}\,(\Sigma ')^2~.\label{12} 
\eea
Although $W$ has the parity, $\phi\to -\phi$
and $\Sigma ' \to -\Sigma '$, 
it is not the most general potential, but, as we already
mentioned, this does not contradict the philosophy of 
the coupling unification by the RGI method.

\begin{figure}[t]
  \begin{center}
    \leavevmode
    \rotate[r]{
      \mbox{
        \epsfysize=11cm
        \epsffile{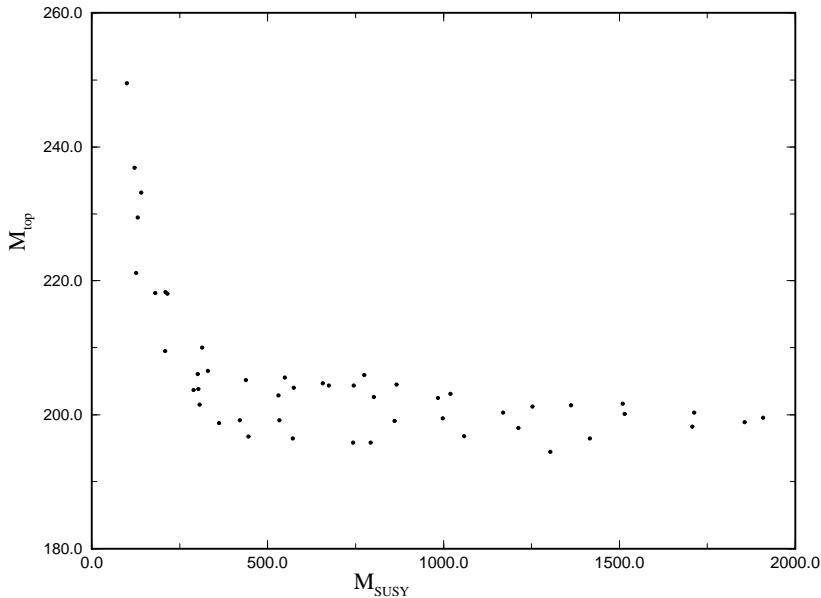} 
        }
      }
  \end{center}
  \caption[\ ]{The values for $M_t$ predicted by the Pati-Salam model
    for different $M_{SUSY}$ scales.  Only the ones with $M_{SUSY}$
    beyond 400 GeV are realistic.} 
  \label{fig:2}
\end{figure}

We denote the gauge couplings of $SU(4)\times SU(2)_{R}\times
SU(2)_{L}$
by $\alpha_{4}~,~\alpha_{2R}$ and $\alpha_{2L}$,
respectively. The gauge coupling for $U(1)_{Y}$, $\alpha_1$, normalized
in the usual GUT inspired manner, is given by
$1/\alpha_{1} ~=~2/5\alpha_{4}+
3/5 \alpha_{2R}~$.
In principle, the primary coupling can be any one of the couplings.
But it is more convenient to choose a gauge coupling as the primary
one because the one-loop $\beta$ functions for a gauge coupling
depends only on its own gauge coupling. For the present model,
we use $\alpha_{2L}$ as the primary one.
Since the gauge sector for the one-loop $\beta$ functions is closed,
the solutions of the fixed point equations (\ref{fixpt}) are 
independent on the Yukawa and Higgs couplings. One easily obtains
$
\rho_{4}^{(1)} =8/9~,~\rho_{2R}^{(1)}~=~4/5$,
so that the
RGI relations (\ref{algeq}) at the one-loop level become
\bea
\tilde{\alpha}_{4} &=&\frac{\alpha_4}{\alpha_{2L}}~=~
\frac{8}{9}~,~\tilde{\alpha}_{1} ~=~\frac{\alpha_1}{\alpha_{2L}}~=~
\frac{5}{6}~.
\label{13}
\eea

The solutions in the Yukawa-Higgs sector strongly
depend on the result of the gauge sector. After slightly involved
algebraic computations, one finds that 
most predictive solutions contain at least
three vanishing $\rho_{i}^{(1)}$'s.  
Out of these solutions, there are two that
 exhibit the most predictive
power and moreover they satisfy
the neutrino mass relation
$m_{\nu_{\tau}}~>~m_{\nu_{\mu}}~,~
m_{\nu_{e}}$. 
For the first solution we have $\rho_{1\phi}^{(1)}=
\rho_{2\phi}^{(1)}=
\rho_{\Sigma}^{(1)}=0$, while for the second solution, 
$  \rho_{1\phi}^{(1)}=
\rho_{2\phi}^{(1)}=
\rho_{G}^{(1)}=0 $,
 and one finds that for the cases above the power series solutions
(\ref{algeq}) take the form 
\bea
\tilde{\alpha}_{GJ} &\simeq &
\left\{
\begin{array}{l} 1.67 - 0.05 \tilde{\alpha}_{1\phi}
+ 
0.004 \tilde{\alpha}_{2\phi}
 - 0.90\tilde{\alpha}_{\Sigma}+\cdots \\
 2.20 - 0.08 \tilde{\alpha}_{2\phi}
 - 0.05\tilde{\alpha}_{G}+\cdots 
\end{array} \right. ~~,\nn\\
\tilde{\alpha}_{33} &\simeq&\left\{
\begin{array}{l}  3.33 + 0.05 \tilde{\alpha}_{1\phi} 
+ 
0.21 \tilde{\alpha}_{2\phi}-0.02 \tilde{\alpha}_{\Sigma}+ \cdots 
\\3.40 + 0.05 \tilde{\alpha}_{1\phi} 
-1.63 \tilde{\alpha}_{2\phi}- 0.001 \tilde{\alpha}_{G}+ 
\cdots \end{array} \right. ~~,\nn\\
\tilde{\alpha}_{3\phi} &\simeq&
\left\{
\begin{array}{l}  1.43 -0.58 \tilde{\alpha}_{1\phi} 
- 
1.43 \tilde{\alpha}_{2\phi}-0.03 \tilde{\alpha}_{\Sigma}+ 
\cdots\\
 0.88 -0.48 \tilde{\alpha}_{1\phi} 
+8.83 \tilde{\alpha}_{2\phi}+ 0.01 \tilde{\alpha}_{G}+ 
\cdots\end{array} \right. ~~,\nn\\
\tilde{\alpha}_{H} &\simeq& \left\{
\begin{array}{l}
 1.08 -0.03 \tilde{\alpha}_{1\phi} 
+0.10 \tilde{\alpha}_{2\phi}- 0.07 \tilde{\alpha}_{\Sigma}+ 
\cdots\nn\\
2.51 -0.04 \tilde{\alpha}_{1\phi} 
-1.68 \tilde{\alpha}_{2\phi}- 0.12 \tilde{\alpha}_{G}+ 
\cdots\end{array} \right. ~~,~~\\
\tilde{\alpha}_{\Sigma} &\simeq& \left\{
\begin{array}{l}
---\\
0.40 +0.01 \tilde{\alpha}_{1\phi} 
-0.45 \tilde{\alpha}_{2\phi}-0.10 \tilde{\alpha}_{G}+ 
\cdots \end{array} \right. ~,\nn\\  
\tilde{\alpha}_{\Sigma '} &\simeq& \left\{
\begin{array}{ll}
4.91 - 0.001 \tilde{\alpha}_{1\phi} 
-0.03 \tilde{\alpha}_{2\phi}- 0.46 \tilde{\alpha}_{\Sigma}+ 
\cdots
\\8.30 + 0.01 \tilde{\alpha}_{1\phi} 
+1.72 \tilde{\alpha}_{2\phi}- 0.36 \tilde{\alpha}_{G}+ 
\cdots \end{array} \right. ~~,  \nn\\
\tilde{\alpha}_{G} &\simeq& \left\{
\begin{array}{ll}
5.59 + 0.02 \tilde{\alpha}_{1\phi} 
-0.04 \tilde{\alpha}_{2\phi}- 1.33 \tilde{\alpha}_{\Sigma}+ 
\cdots
\\--- \end{array} \right.   ~ ~.\label{14}
\eea
We have assumed that the Yukawa couplings $g_{IJ}$ except for
$g_{33}$ vanish. They can be included into RGI relations
as small 
perturbations,
but their numerical effects
will be rather small.

 The number $ N_{H}$ of the Higgses lighter
than $M_{SUSY}$ could vary from one to four while the number of
those to be taken into account above $M_{SUSY}$ is fixed at four.
We have assumed here that $N_{H}=1$. The dependence of the top mass on
$M_{SUSY}$ in this model is shown in Figure \ref{fig:2}.

\section{Asymptotically Non-Free SO(10) Model}

We will show in this section a model based on $SO(10)$ in which also
the reduction of couplings can be applied \cite{kmtz2}.

We denote the hermitean
$SO(10)$-gamma matrices  by
$\Gamma_{\alpha}~,~\alpha=1,\cdots,10$.
The charge conjugation matrix $C$ satisfies
$C = C^{-1}~,~C^{-1}\,\Gamma_{\alpha}^{T}\,C ~=~
-\Gamma_{\alpha}$, and 
the $\Gamma_{11}$ is defined as
$\Gamma_{11} \equiv (-i)^5 \,\Pi_{\alpha =1}^{10}
\Gamma_{\alpha} ~~ \mbox{with} ~~(\Gamma_{11})^2 ~=~1$.
The chiral projection operators are given by
${\cal P}_{\pm} = \frac{1}{2}(\,1\pm \Gamma_{11})$.

In $SO(10)$ GUTs \cite{fritzsch1,mohapatra1},
three generations of quarks and leptons are
 accommodated by 
three chiral supermultiplets in
${\bf 16}$ which we
denote by 
\be
\Psi^{I}({\bf 16})~~\mbox{with}~~{\cal P}_{+}\,
\Psi^{I}~=~\Psi^{I}~,
\ee
where $I$ runs over the three generations
and the spinor index is suppressed.
To break $SO(10)$ down to $SU(3)_{\rm C}
\times SU(2)_{\rm L} \times U(1)_{\rm Y}$, we use
the following set of chiral superfields:
\be
S_{\{\alpha\beta\}}({\bf 54})~,~
A_{[\alpha\beta]}({\bf 45})~,~
\phi({\bf 16})~,~\overline{\phi}({\overline{\bf 16}})~.
\ee
The two $SU(2)_{\rm L}$ doublets which are responsible for
the spontaneous symmetry breaking (SSB) of 
$SU(2)_{\rm L} \times U(1)_{\rm Y}$ down to $U(1)_{\rm EM}$
are contained in
$ H_{\alpha}({\bf 10})$.
We further introduce a singlet $\varphi$ which after
the  SSB of $SO(10)$ will
mix with the right-handed neutrinos 
so that they will become superheavy.

The superpotential of  the model is 
given by 
\bea
W &=& W_{Y}+W_{SB}+ W_{HS}+ W_{NM}+W_{M}~,
\eea
where
\bea
W_{Y} &=&\frac{1}{2}\sum_{I,J=1}^{3}g_{IJ}\,\Psi^{I}\,C\Gamma_{\alpha} 
\,\Psi^{J}~H_{\alpha}~,\nn\\
W_{SB} &=&\frac{g_{\phi}}{2}
\,\overline{\phi}\,
\Gamma_{[\alpha\beta]}\,\phi~A_{[\alpha\beta]}
+\frac{g_{S}}{3!}\,
\mbox{Tr}~S^3+
\frac{g_{A}}{2}\,\mbox{Tr}~A^2\,S~, \\
W_{HS} &=&
\frac{g_{HS}}{2}\,\
H_{\alpha}\,S_{ \{\alpha\beta \} }\,H_{\beta}~,~
W_{NM}^{I} ~=~ \sum_{I=1}^{3}\,g_{INM}\,\Psi^I\,
\overline{\phi}\,\varphi~,\nn\\
W_{M}&=&\frac{m_{H}}{2}\,H^2+m_{\varphi}\,\varphi^2
+m_{\phi}\,\overline{\phi} \phi+\frac{m_{S}}{2}\,S^2+
\frac{m_{A}}{2}\,A^2~ ,\nn 
\eea
and $\Gamma_{[\alpha\beta]}=i
 (\Gamma_{\alpha}\Gamma_{\beta}-
\Gamma_{\beta}\Gamma_{\alpha})/2$.
As in the case of the $SU(5)$ minimal model, the superpotential is not
the most general one, but 
this does not contradict the philosophy of 
the coupling unification by the reduction 
method.
$W_{SB}$  is responsible for
the SSB of $SO(10)$ down to
$SU(3)_{C}\times SU(2)_{W}\times U(1)_{Y}$, 
and this can be achieved without breaking supersymmetry,
while $W_{HS}$ is responsible for
the triplet-doublet splitting of $H$.
The right-handed neutrinos 
obtain a superheavy mass through $W_{NM}$ after the
SSB, and the Yukawa couplings 
for the leptons and quarks are contained in $W_{Y}$.
We assume  that
 there exists a choice of soft supersymmetry breaking terms so that
all the vacuum expectation values necessary for the desired SSB 
corresponds to the minimum of
the potential.

Given the supermultiplet content and the superpotential $W$,
we  can  compute the $\beta$ functions of the model. 
The gauge coupling of $SO(10)$ is denoted by $g$, and 
our  normalization of the $\beta$ functions
is as usual, i.e., 
$d g_{i}/d \ln \mu ~=~
\beta^{(1)}_{i}/16 \pi^2+O(g^5)$,
where $\mu$ is the renormalization
scale. We find: 
\bea
  \beta^{(1)}_{g} &=&7\,g^3~,\nn\\
  \beta^{(1)}_{g_T} &=& g_T\,(\,14 |g_T|^2+\frac{27}{5}|g_{HS}|^2+
|g_{3NM}|^2-\frac{63}{2}g^2\,)~,\nn\\
  \beta^{(1)}_{g_{\phi}} &=& g_{\phi}(\,53 |g_{\phi|^2}+
\frac{48}{5}|g_{A}|^2+\frac{1}{2}|g_{1NM}|^2+\frac{1}{2}|g_{2NM}|^2+
\frac{1}{2}|g_{3NM}|^2-\frac{77}{2}g^2\,),\nn\\
\beta^{(1)}_{S} &=&g_{S}(\,\frac{84}{5}|g_{S}|^2
+12|g_{A}|^2+\frac{3}{2}|g_{HS}|^2-60 g^2\,)~,\nn\\
\beta^{(1)}_{A} &=&
g_{A}(\,16|g_{\phi}|^2+
\frac{28}{5}|g_{S}|^2+\frac{116}{5}|g_{A}|^2+\frac{1}{2}|g_{HS}|^2 -52
g^2\,)~,\nn\\
  \beta^{(1)}_{HS} &=&g_{HS}(\,8|g_{T}|^2+
\frac{28}{5}|g_{S}|^2+4|g_{A}|^2+\frac{113}{10}|g_{HS}|^2
-38 g^2\,) ~,\\
\beta^{(1)}_{1NM} &=& g_{1NM}(\,\frac{45}{2}|g_{\phi}|^2+
9 |g_{1NM}|^2+\frac{17}{2}|g_{2NM}|^2+
\frac{17}{2}|g_{3NM}|^2-\frac{45}{2}g^2\,)~,\nn\\
 \beta^{(1)}_{2NM} &=& g_{2NM}(\,\frac{45}{2}|g_{\phi}|^2+
\frac{17}{2} |g_{1NM}|^2+9 |g_{2NM}|^2+
\frac{17}{2}|g_{3NM}|^2-\frac{45}{2}g^2\,)~,\nn\\
\beta^{(1)}_{3NM} &=& g_{3NM}(\,5 |g_T|^2+\frac{45}{2}|g_{\phi}|^2+
\frac{17}{2} |g_{1NM}|^2+\frac{17}{2}|g_{2NM}|^2\nn\\
&&+9|g_{3NM}|^2-\frac{45}{2}g^2\,)~.\nn
\eea
We have assumed that the Yukawa couplings $g_{IJ}$ except for
$g_T \equiv g_{33}$ vanish. They can be included as small
perturbations. Needless to say that the
soft susy breaking terms do not alter the $\beta$ functions
above.

 \begin{figure}[t]
           \epsfxsize= 11 cm   
           \centerline{\epsffile{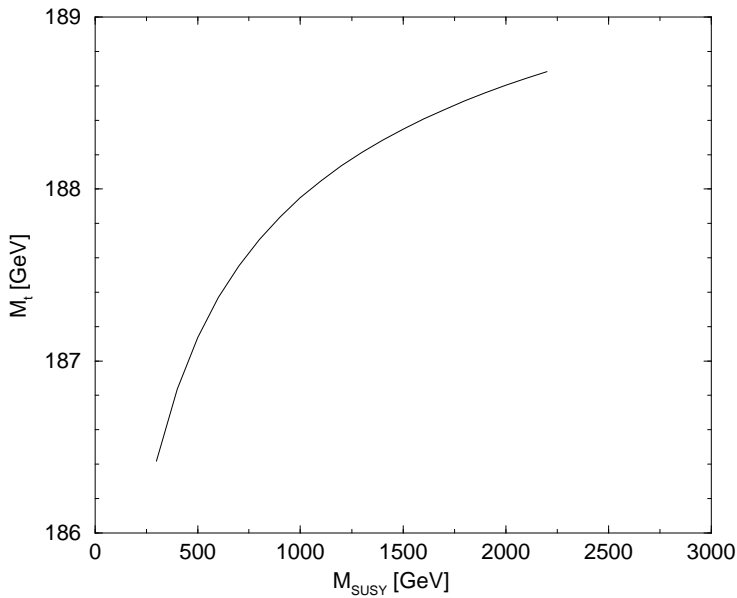}}
        \caption{$M_t$ prediction versus 
$M_{\rm SUSY}$ for $\tilde{\alpha}_{T}=2.717$.}
        \label{fig:sotms}
        \end{figure}

We find that 
there exist two independent
solutions,
$A$ and $B$, that have the most predictive 
power, where we have chosen the $SO(10)$ gauge coupling as
the primary coupling:
\bea
\rho_{T} &= &\left\{
\begin{array}{ll} 163/60 &\simeq 2.717 \\
0 &  \end{array} \right. ~~,~~
\rho_{\phi} ~= ~\left\{
\begin{array}{ll} 5351/9180 &\simeq 0.583 \\
1589/2727 &\simeq 0.583 \end{array} \right. ~,\nn\\  
\rho_{S} &= &\left\{
\begin{array}{ll} 152335/51408 &\simeq 2.963 \\
850135/305424 &\simeq 2.783 \end{array} \right. ,
\rho_{A}=\left\{
\begin{array}{ll} 31373/22032 &\simeq 1.424 \\
186415/130896 &\simeq 1.424 \end{array} \right.,\nn\\ 
\rho_{HS}&= &\left\{
\begin{array}{ll} 7/81 & \simeq 0.086 \\
170/81 &\simeq 2.099 \end{array} \right. ~,~
\rho_{1NM} = \rho_{2NM} =\left\{
\begin{array}{ll} 191/204 &\simeq 0.936 \\
191/303 &\simeq 0.630 \end{array} \right. ~~,\nn\\
\rho_{3NM}&= &\left\{
\begin{array}{ll} 0 &  \\
191/303 &\simeq 0.630 \end{array} \right. ~
~~\mbox{for}~~\left\{\begin{array}{l} A   \\
B\end{array} \right.~.
\eea
Clearly, the solution B has less predictive power because
$\rho_T =0$. So, we consider below only the solution A,
in which the coupling $\alpha_{3NM}$ should be
regarded as a small perturbation because $\rho_{3NM}=0$.

Given this solution it is possible to show, as in the case of
$SU(5)$, that the $\rho$'s can be uniquely computed in any finite order
in perturbation theory.

The corrections to the reduced couplings coming from
the small perturbations up to and including terms of
$O(\tilde{\alpha}_{3NM}^2$):
\bea
\label{alfas-so10}
\tilde{\alpha}_T &=&
(\,163/60 - 0.108\cdots  \tilde{\alpha}_{3NM} + 
0.482 \cdots  \tilde{\alpha}_{3NM}^2+\cdots\,)
+\cdots~,\nn\\
\tilde{\alpha}_{\phi} &=&
(\,5351/9180 + 0.316\cdots  \tilde{\alpha}_{3NM} +
0.857\cdots  \tilde{\alpha}_{3NM}^2+\cdots\,)
+\cdots~,\nn\\
\tilde{\alpha}_{S} &=&
(\,152335/51408 + 0.573\cdots  \tilde{\alpha}_{3NM} 
+ 5.7504\cdots  \tilde{\alpha}_{3NM}^2+\cdots\,)
+\cdots~,\nn\\
\tilde{\alpha}_{A} &=&
(\,31373/22032 - 0.591\cdots  \tilde{\alpha}_{3NM} 
- 4.832\cdots  \tilde{\alpha}_{3NM}^2+\cdots\,)
+\cdots~,\nn\\
\tilde{\alpha}_{HS} &=&
(7/81 - 0.00017\cdots  \tilde{\alpha}_{3NM} 
+ 0.056\cdots  \tilde{\alpha}_{3NM}^2+\cdots\,)
+\cdots~,\\
\tilde{\alpha}_{1NM}&=&\tilde{\alpha}_{2NM}=
(\,191/204 - 4.473\cdots  \tilde{\alpha}_{3NM} 
+ 2.831\cdots  \tilde{\alpha}_{3NM}^2+\cdots\,)
+\cdots,\nn
\eea
where $\cdots$ indicates higher order terms which
can be uniquely computed. 
In the partially reduced theory defined above,
we have two independent couplings,
$\alpha$ and $\alpha_{3NM}$ (along with the
Yukawa couplings $\alpha_{IJ}~,~ I,J\neq T$).

At the one-loop level,
Eq.~(\ref{alfas-so10}) defines a line parametrized by $\tilde{\alpha}_{3NM}$
in the $7$ dimensional space of couplings.
A numerical analysis shows  that this line blows up
in the direction of $\tilde{\alpha}_{S}$ at
a finite value of $\tilde{\alpha}_{3NM}$ \cite{kmtz2}.
So if we require $\tilde{\alpha}_{S}$ to remain within
the perturbative regime (i.e., $g_S \leq 2$,
which means $ \tilde{\alpha}_{S} \leq 8$ because 
$\alpha_{\rm GUT} \sim 0.04$),
the $\tilde{\alpha}_{3NM}$ should be restricted to be below
$\sim 0.067$. As a consequence, the value of
$\tilde{\alpha}_{T}$ is also bounded
\be
2.714 \leq \tilde{\alpha}_{T} \le 2.736~.
\ee
This defines 
GYU boundary conditions holding 
at the unification scale $M_{\rm GUT}$ in addition to the 
group theoretic one,
$\alpha_{T}=\alpha_{t} ~=~\alpha_{b} ~=~\alpha_{\tau}$.
The value of $ \tilde{\alpha}_{T}$ is practically fixed so that 
we may assume that
$\tilde{\alpha}_{T}=163/60 \simeq 2.72$,
which is the unperturbed value.

 Figure \ref{fig:sotms} shows the prediction for $M_t$ in this model
 for different values of the supersymmetry 
breaking scale $M_{susy}$.  It is worth noticing that the value for
$M_t$ predicted is below its infrared value ($M_{top-IR} \sim 189 GeV$)
\cite{kmtz2}, but it is
slightly above the recent experimental values (\ref{mtop-exp}).

\section{Conclusions}

As a natural extension of the unification of gauge couplings provided by 
all GUTs and the unification of Yukawa couplings, we
have introduced the idea of Gauge-Yukawa
Unification. GYU is a functional relationship among the gauge and
Yukawa couplings provided by some principle.  In our studies GYU has
been achieved by applying the principles of reduction of couplings
and finiteness. 
The consequence of GYU is that 
in the lowest order in perturbation theory
 the gauge and Yukawa couplings above  $M_{\rm GUT}$
are related  in the form
\be
g_i  = \kappa_i \,g_{\rm GUT}~,~i=1,2,3,e,\cdots,\tau,b,t~,
\label{bdry}
\ee 
where $g_i~(i=1,\cdots,t)$ stand for the gauge 
and Yukawa couplings, $g_{\rm GUT}$ is the unified coupling,
and
we have neglected  the Cabibbo-Kobayashi-Maskawa mixing 
of the quarks.
 So, Eq.~(\ref{bdry}) exhibits a set of boundary conditions on the 
the renormalization group evolution for the effective theory
below $M_{\rm GUT}$, which we have assumed to 
be the MSSM. 
We have shown \cite{kmoz,kmoz2} that it is
possible to construct some 
supersymmetric GUTs with GYU in the 
third generation that can
 predict the bottom and top
quark masses in accordance with the recent experimental data
\cite{top}. 
This means that the top-bottom hierarchy 
could be
explained in these models,
 in a similar way as 
the hierarchy of the gauge couplings of the SM
can be explained if one assumes  the existence of a unifying
gauge symmetry at $M_{\rm GUT}$. 

 \begin{figure}[t]
  \begin{center}
    \leavevmode
    \rotate[r]{
      \mbox{
        \epsfysize=11cm
        \epsffile{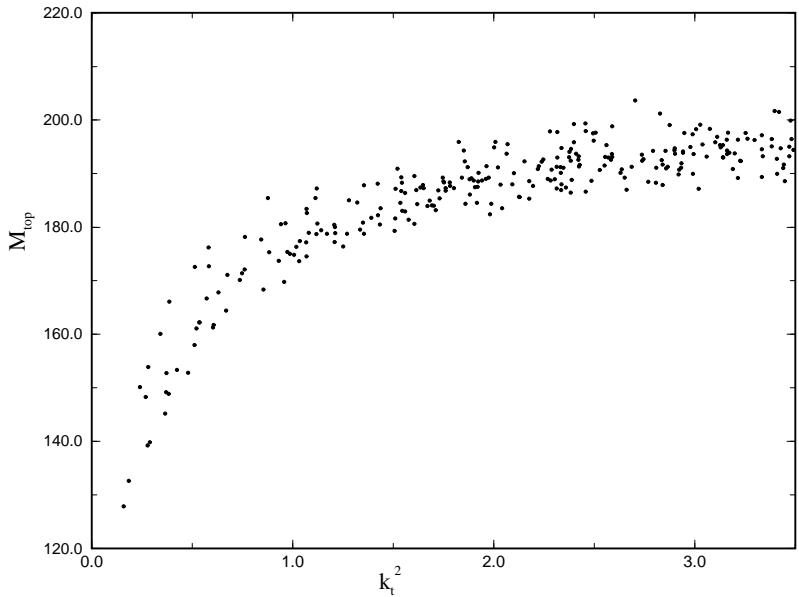}
        }
      }
  \end{center}
  \caption[\ ]{The dependence of the top mass $M_t$ with $k^2_t$, at
    fixed $M_{SUSY}=500$ GeV.
    As we can see, after $k^2_t\sim 2.0$ the top mass goes to its infrared
    fixed point value.}
  \label{fig:3}
\end{figure}

It is clear that the GYU scenario  is the most predictive scheme as far as
the mass of the top quark is concerned.
It may be worth recalling the predictions for $M_t$
of ordinary GUTs, in particular of supersymmetric $SU(5)$ and
$SO(10)$.  The MSSM with $SU(5)$ Yukawa boundary unification allows
$M_t$ to be anywhere in the interval between 100-200 GeV 
for varying $\tan \beta$, which is now a free parameter.  Similarly,
the MSSM with $SO(10)$ Yukawa 
boundary conditions, {\em i.e.} $t-b-\tau$ Yukawa Unification gives
$M_t$ in the interval 160-200 GeV. 
We have analyzed \cite{kmoz2} the infrared quasi-fixed-point behaviour of
the $M_t$ prediction in some detail. In particular we have seen that
the {\em infrared value} for large $\tan \beta$ depends on  $\tan
\beta$  and its lowest value is $\sim 188$ GeV.  Comparing this with
the experimental value ({\ref{mtop-exp}) we may conclude
that the present data on $M_t$ cannot be explained from the infrared
quasi-fixed-point behaviour alone (see Figure 4).

Clearly, to exclude or verify different GYU models,
 the experimental as well as theoretical uncertainties
have to be further reduced.
One of the largest theoretical uncertainties 
 in FUT  results
from the not-yet-calculated threshold effects 
of the superheavy particles.
Since the structure of  the superheavy 
particles is basically fixed,
 it will be possible to
bring these threshold effects under control,
which will  reduce the uncertainty of 
the $M_t$ prediction.
 We have been regarding $\delta^{\rm MSSM} M_t$ 
as unknown because we do not have 
sufficient information on the superpartner spectra.
Recently, however, we have demonstrated \cite{kmz-pert} how to extend
 the principle of reduction of couplings in a way as to include the
 dimensionfull parameters.  As a result, it is in principle possible
 to predict the superpartner spectra as well as the rest of the
 massive parameters of a theory. 

\section*{Acknowledgements}

One of us (G.Z.) would like to thank the Organizing Committees of the
Schools for their warm hospitality.


\begin{thebibliography}{100}

\bibitem{pati1} J.C. Pati and A. Salam, {\sl Phys. Rev. Lett.} {\bf 31}
                  (1973) 661.
\bibitem{gut1} H. Georgi and S.L. Glashow,
             {\sl Phys. Rev. Lett.} {\bf 32} (1974) 438;
        H. Georgi, H. Quinn, S. Weinberg, {\sl Phys. Rev. Lett.}
        {\bf 33} (1974) 451.

\bibitem{fritzsch1} H. Fritzsch and P. Minkowski,
               {\sl Ann. Phys.} {\bf 93} (1975) 193;
              H. Georgi, in {\sl Particles and Fields -- 1974},
              ed. C.E. Carlson, (American Institute of Physics, New York).

\bi{begn} A. Buras, J. Ellis, M.K. Gaillard and D. Nanopoulos,
           {\sl Nucl. Phys.} {\bf B135} (1978) 66.

\bi{abf} U. Amaldi, W. de Boer and H. F\"urstenau, {\sl {\sl Phys. Lett.}} 
                {\bf B260} (1991) 447.

\bi{depe} R. Decker and J. Pestieau,{\sl Lett. Nuovo Cim.} {\bf 29}
   (1980) 560, M. Veltman, {\sl Acta Phys. Pol.} {\bf B12} (1981) 437.

\bi{fgp-prd90} S. Ferrara, L. Girardello and F. Palumbo, {\sl
  Phys. Rev.} {\bf D20} (1979) 403.

\bi{pr}B. Pendleton and G.G Ross, {\sl Phys. Lett.} {\bf B98}
(1981) 291.

\bi{zim-pr} W. Zimmermann, {\sl Phys. Lett.} {\bf B308} (1993) 117.

\bi{hill}C.T. Hill, {\sl Phys. Rev.} {\bf D24} (1981) 691;  \newline
W.A. Bardeen, C.T. Hill and M. Lindner, {\sl Phys. Rev.} {\bf D411}
(1990) 1647. 

\bibitem{kmz} D. Kapetanakis, M. Mondrag{\' o}n and
    G. Zoupanos, {\sl Zeit. f. Phys.} {\bf C60} (1993) 181;
    M. Mondrag{\' o}n and G. Zoupanos, {\sl Nucl. Phys.} {\bf B}
    (Proc. Suppl) {\bf 37C} (1995) 98.

\bi{mondragon2} J. Kubo, M. Mondrag\'on and G. Zoupanos,
{\sl Nucl. Phys.} {\bf B424} (1994) 291.

\bi{kmtz} J. Kubo, M. Mondrag{\' o}n, N.D. Tracas and
G. Zoupanos, {\sl Phys. Lett.} {\bf B342} (1995) 155.

\bi{kmtz2} J. Kubo, M. Mondrag{\' o}n, S. Shoda and
G. Zoupanos, {Nucl. Phys.} {\bf B469} (1996) 3.

\bi{kmoz} J. Kubo, M. Mondrag{\' o}n, M. Olechowski and
G. Zoupanos, {\em Gauge-Yukawa Unification and the Top-Bottom
  Hierarchy},  Proc. of the {\em
  Int. Europhysics Conf. on HEP}, Brussels 1995.

\bi{kmoz2} J. Kubo, M. Mondrag{\' o}n, M. Olechowski and
G. Zoupanos, {\em Testing Gauge-Yukawa-Unified
Model by $M_t$}, hep-ph/9512435, to be published in {\sl Nucl. Phys.} {\bf
B}.

\bi{cheng1} T.P. Cheng, E. Eichten and L.F. Li, {\sl Phys. Rev.}
   {\bf D9} (1974) 2259; N.P. Chang, {\sl Phys. Rev.} {\bf D10} (1974) 2706;
   E. Ma, Phys. Rev {\bf D17} (1978) 623; {\sl ibid} {\bf D31} (1985) 1143. 

\bibitem{zim1} W. Zimmermann, {\sl Com. Math. Phys.}
              {\bf 97} (1985) 211;
 R. Oehme and W. Zimmermann {\sl Com. Math. Phys.}
              {\bf 97} (1985) 569; R. Oehme, K. Sibold and W. Zimmermann,
                {\sl Phys. Lett.} {\bf B147} (1984) 117;
              {\bf B153} (1985) 142; 
R. Oehme, {\sl Prog. Theor. Phys. Suppl.}
              {\bf 86} (1986) 215


\bibitem{kubo1} J. Kubo, K. Sibold and W. Zimmermann,
             {\sl Nucl. Phys.} {\bf B259} (1985) 331; 
{\sl Phys. Lett.} {\bf B200} (1989) 185.

\bibitem{kubo3} J. Kubo, {\sl Phys. Lett.} {\bf B262} (1991) 472.

\bibitem{kubo2} J. Kubo, K. Sibold and W. Zimmermann, {\sl
    Phys. Lett.} {\bf B200} (1989) 185.

\bibitem{PW} A.J. Parkes and P.C. West, {\sl Phys. Lett.} 
{\bf B138} (1984) 99;
             {\sl Nucl. Phys.} {\bf B256} (1985) 340;
             P. West, {\sl Phys. Lett.} {\bf B137} (1984) 371;
             D.R.T. Jones and A.J. Parkes, {\sl Phys. Lett.} {\bf B160} (1985)
             267;
             D.R.T. Jones and L. Mezinescu, {\sl Phys. Lett.} {\bf B136} (1984)
             242; {\bf B138} (1984) 293;
             A.J.~Parkes, {\sl Phys. Lett.} {\bf B156} (1985) 73.

\bibitem{HPS} S. Hamidi, J. Patera and J.H. Schwarz,
              {\sl Phys. Lett.} {\bf B141} (1984) 349;
           X.D. Jiang and X.J. Zhou,
        {\sl Phys. Lett.} {\bf B197} (1987) 156; {\bf B216} (1985) 160.

\bi{soft}
D.R.T. Jones, L. Mezincescu and Y.-P. Yao, {\sl Phys. Lett.}
{\bf B148} (1984) 317.

\bi{jj}I. Jack and D.R.T. Jones, {\sl Phys.Lett.} {\bf B333} (1994) 372.

\bi{strassler} R.G. Leigh and M.J. Strassler, {\sl Nucl. Phys.}  {\bf
  B447} (1995) 95. 

\bibitem{model1} S. Hamidi and J.H.~Schwarz,
              {\sl Phys. Lett.} {\bf B147} (1984) 301;
                D.R.T. Jones and S. Raby,
              {\sl Phys. Lett.} {\bf B143} (1984) 137;
              J.E. Bjorkman,  D.R.T. Jones and S. Raby
              {\sl Nucl. Phys.} {\bf B259} (1985) 503.  

\bibitem{model}J. Le\'on et al,
              {\sl Phys. Lett.} {\bf B156} (1985) 66.


\bi{zimmermann3}W. Zimmermann, {\sl Phys. Lett.} {\bf B311} (1993)
249.

\bi{sakai1} S. Dimopoulos and H. Georgi, {\sl Nucl. Phys.} {\bf B193}
(1981) 150;
N. Sakai, {\sl Zeit. f. Phys.} {\bf C11} (1981) 153.

\bi{raifer} L. O'Raifeartaigh, {\sl Nucl. Phys.} {\bf B96} (1975) 331.

\bi{fayet} P. Fayet and J. Iliopoulos, {\sl Phys. Lett.} {\bf B51}
(1974) 461. 

\bibitem{nonre} J. Wess and B. Zumino, Phys. Phys. {\bf B49} 52;
J. Iliopoulos and B. Zumino, {\sl Nucl. Phys.} {\bf B76} (1974) 310;
S. Ferrara, J. Iliopoulos and B. Zumino,
{\sl Nucl. Phys.} {\bf B77} (1974) 413;
K. Fujikawa and W. Lang, {\sl Nucl. Phys.} {\bf B88} (1975)
61.

\bibitem{LPS} C. Lucchesi, O. Piguet and K. Sibold,
                {\sl Helv. Phys. Acta} {\bf 61} (1988) 321; {\sl
                Phys. Lett.} {\bf B201} (1988) 241.

\bibitem{pisi} O. Piguet and K. Sibold, {\sl Int. Journ. Mod. Phys.} {\bf A1}
               (1986) 913; {\sl Phys. Lett.} {\bf B177} (1986) 373;
               R. Ensign and K.T. Mahanthappa, {\sl Phys. Rev.} {\bf
               D36} (1987) 3148.

\bibitem{LZ} C. Lucchesi and G. Zoupanos,
{\sl All Order Finiteness in 
$N=1$ SYM Theories: Criteria and Applications},
hep-ph/9604216.

\bi{piguet} O. Piguet, {\em Supersymmetry, Ultraviolet Finiteness and
  Grand Unification}, hep-th/9606045.

\bi{ag-npb243} L. Alvarez-Gaum\'e and P. Ginsparg, {\sl Nucl. Phys.}
{\bf B243} (1984) 449; W.A. Bardeen and B. Zumino, {\sl
  Nucl. Phys.}{\bf B243} (1984) 421; B. Zumino, Y. Wu and A. Zee, {\sl
  Nucl. Phys.} {\bf B439} (1984) 477.

\bi{ab-theo} S.L. Adler and W.A. Bardeen, {\sl Phys. Rev.} {\bf 182}
(1969) 1517. 

\bi{fz-npb87} S. Ferrara and B. Zumino, {\sl Nucl. Phys.} {\bf B87}
(1975) 207. 

\bi{pisi-npb196}  O. Piguet and K. Sibold, {\sl Nucl. Phys.} {\bf
  B196} (1982) 428; {\bf B196} (1982) 447.

\bi{pisi-book} O. Piguet and K. Sibold, {\sl Renormalized
  Supersymmetry}, Birkh\"auser Boston, 1986.  

\bi{proton}
N. Deshpande, Xiao-Gang, He and E. Keith, {\sl Phys. Lett.}
{\bf B332} (1994) 88.


 
\bi{barger}H. Arason {\em et al}., Phys. Rev.
{\bf D46} (1992) 3945;
 V. Barger, M.S. Berger and P. Ohmann,
Phys. Rev. {\bf D47} (1993) 1093, and references therein.

\bi{hall1}
L. Hall, R. Rattazzi and U. Sarid, Phys. Rev. {\bf D50}
(1994) 7048;
 M. Carena, M. Olechowski, S. Pokorski and C.E.M. Wagner,
Nucl. Phys. {\bf B426} (1994) 269.

\bi{wright1}
B.D. Wright, {\em Yukawa Coupling Thresholds:
Application to the MSSM and the Minimal Supersymmetric SU(5) GUT},
University of Wisconsin-Madison report, MAD/PH/812 (hep-ph/9404217).

\bi{pdg}Particle Data Group, L. Montanet {\em et al}.,
Phys. Rev. {\bf D50} (1994) 1173.

\bi{pokorski1}
P.H. Chankowski, Z. Pluciennik and S. Pokorski,
Nucl. Phys. {\bf B439} (1995) 23.

\bi{langacker1}
P. Langacker and N. Polonsky, Phys. Rev. {\bf D47} (1993) 4028.

\bi{borzumati1} F.M. Borzumati, M. Olechowski and S. Pokorski,
Phys. Lett. {\bf B349} (1995) 311;
H. Murayama, M. Olechowski and S. Pokorski,
{\em Viable $t-b-\tau$ Yukawa Unification in SUSY $SO(10)$},
MPI preprint No. MPI-PhT/95-100, hep-ph/9510327.

\bi{polonsky1}
D. Pierce, in {\em Proc. of SUSY 94}, Ann Arbor, Michigan,
1994, eds. C. Kolda and J. Wells, p.418;
A. Donini, {\em One-loop Corrections to the Top,
Stop and Gluino Masses in the MSSM},
CERN report, CERN-TH-95-287 (hep-ph/9511289);
J. Feng, N. Polonsky and S.
Thomas,  {\em The Light Higgsino-Gaugino Window},
University of
Munich report, LMU-TPW-95-18.

\bi{hill1}
W.A. Bardeen, M. Carena, S. Pokorski and C.E.M. Wagner,
{\sl Phys. Lett.} {\bf B320} (1994) 110;
M. Carena, M. Olechowski, S. Pokorski and C.E.M. Wagner,
{\sl Nucl. Phys.} {\bf B419} (1994) 213.

\bi{threshold}J. Hisano, H. Murayama and T. Yanagida,
Phys. Rev. Lett. {\bf 69} (1993) 1992;
J. Ellis, S. Kelley and D. V. Nanopoulos,
Nucl. Phys. {\bf B373} (1992) 55;
Y. Yamada, Z. Phys. {\bf C60} (1993) 83.

\bi{top}
J. Lys, The CDF Collaboration, FERMILAB-CONF-96/409-E,
Proceedings of the {\em 28th International Conference on High Energy
  Physics (ICHEP'96)};
S. Protopopescu, D0 Collaboration, Proceedings of the {\em 28th
  International Conference on High Energy 
  Physics (ICHEP'96)}, Warsaw, Poland, July 25-31, 1996.


\bi{KYIS95}  J. Kubo, M. Mondrag\'on and G. Zoupanos, 
 {\em The Top-Bottom  Hierarchy  
from  Gauge-Yukawa Unification}, to be published in the
Proc. of the {\em International Seminar '95}, 
August 21-25, 1995, Kyoto.

\bi{HMY-npb402} J. Hisano, H. Murayama and T. Yanagida,
{\sl Nucl. Phys.} {\bf B402}  (1993) 46.

\bi{anton1}I. Antoniadis and G.K. Leontaris, {\sl Phys. Lett.} {\bf B216}
(1989) 333;
G. Leontaris and N. Tracas, Z. Phys. {\bf C56} (1992) 479;
{\sl Phys. Lett.} {\bf B291} (1992) 44.

\bi{georgi4}
H. Georgi and C. Jarlskog, {\sl Phys. Lett.} {\bf B86} (1979) 297.

\bi{mohapatra1}
R. N. Mohapatra, {\em Left-Right Symmetric Models
of Weak Interactions} in Proc. of a NATO ASI
on Quarks, Leptons, and Beyond, September 5-16, 1983,
Munich, eds. H.  Fritsch {et al}. (Plenum Press, New York, 1983).

\bi{barger1} V.Barger et al.  {\sl Phys. Lett.} {\bf B313} (1993) 351;
      M. Carena, S. Pokorski and C. Wagner, {\sl Nucl. Phys.} {\bf B406}
      (1993) 59.

\bi{schrempp1}
B. Schrempp and F. Schrempp,
{\sl Phys. Lett.} 
{\bf B299} (1993) 321;
B. Schrempp,
{\sl Phys. Lett.} 
{\bf B344} (1995) 193; B. Schrempp and W. Wimmer, {\em Top quark and
  Higgs boson masses: interplay between infrared and ultraviolet
  physics}, DESY-960109 preprint, hep-ph/9606386.


\bibitem{al}            A.V. Ermushev, D.I. Kazakov and O.V. Tarasov,
                 {\sl Nucl. Phys.} {\bf B281} (1987) 72;
               D.I. Kazakov, {\sl Mod. Phys. Let.} {\bf A2} (1987) 663; 
                 {\sl Phys. Lett.} {\bf B179} (1986) 352;
               D.I. Kazakov and I.N. Kondrashuk, {\sl Low-energy
             predictions of Susy GUTs: minimal versus finite model},
             preprint E2-91-393.

\bi{anton}
I. Antoniadis, C. Kounnas and K. Tamvakis,
Phys. Lett. {\bf B119} (1982) 377;
A. Sch{\" u}ler, S. Sakakibatra and J.G K{\" o}rner,
{\sl Phys. Lett.} {\bf B194} (1987) 125. 

\bi{kmz-pert} J. Kubo, M. Mondrag\'on and G. Zoupanos, {\em
  Perturbative Unification of Soft Susy Breaking Terms},
  Preprint MPI-PhT/96-71, hep-ph/9609218; see also references therein.
  To be published in {\sl Phys. Lett. B}
 
\end{thebibliography}
\end{document}